\documentclass[useAMS,usenatbib]{mn2e}
\usepackage{graphicx}

\def\la{\raise.5ex\hbox{$<$}\kern-.8em\lower 1mm\hbox{$\sim$}}
\def\ma{\raise.5ex\hbox{$>$}\kern-.8em\lower 1mm\hbox{$\sim$}}

\def\Lsol{L$_{\odot}$ }
\def\kms{$\rm km\, s^{-1}$}
\def\cm3{$\rm cm^{-3}$}
\def\Ts{$\rm T_{*}$~}
\def\Vs{$\rm V_{s}$~}
\def\n0{$\rm n_{0}$}
\def\B0{$\rm B_{0}$}
\def\ne{$\rm n_{e}$~}

\def\Te{$\rm T_{e}$}

\def\Td{$\rm T_{d}$}

\def\erg{$\rm erg\, cm^{-2}\, s^{-1}$}
\def\mum{$\mu$m~}
\def\LIR{L$_{IR}$~}
\def\L12{L$_{12\mu m}$~}
\def\F12{F$_{12\mu m}$~}
\def\agr{a$_{gr}$}
\def\Hb{H${\beta}$~}
\def\Ha{H$\alpha$~}

\title[The merger galaxy Arp 220] {The merger Seyfert galaxy  Arp 220.  Line and continuum absorption and emission}   
\author[M. Contini]{ M. Contini 
\\
School of Physics and Astronomy, Tel Aviv University, Tel Aviv
69978, Israel \\
}

\begin{document}

\date{Accepted: Received ; in original form 2010 month day}

\pagerange{\pageref{firstpage}--\pageref{lastpage}} \pubyear{2009}

\maketitle

\label{firstpage}

\begin{abstract}
The line and continuum spectra of the merger galaxy Arp 220 are analysed with the aim of
investigating the ionizing and heating sources.
We refer  to radio, optical, infrared and X-ray spectra.
The results show that  in agreement with other merger galaxies,
the optical lines are emitted from gas photoionised by the AGN  and heated by the
shocks  in the extended NLR. The  infrared lines are better explained by the emission from gas  
close to the starburst.  The starburst dominates the infrared emission. 
[OI] and [CI] lines in the far-infrared are formed in the internal region of extended
clouds and are therefore absorbed, while [CII] lines are emitted from the external edges of outflowing clouds.
The O/H  relative abundances are about solar and N/H are higher than solar by a factor  $\sim$1.5, 
throughout the starburst region,  
while  in the AGN extended NLR the  O/H ratio is  half solar.
A relatively high dust-to-gas ratio is indicated by  modelling the  dust reprocessed radiation peak 
consistently with bremsstrahlung  emitted from  the clouds. 
The observed radio emission is thermal bremsstrahlung,  while synchrotron radiation created 
by the Fermi mechanism at the shock front is  absorbed.

\end{abstract}


\section{Introduction}

We have been investigating
 the physical and chemical characteristics of some  merger Seyfert galaxies
on the basis of the observed spectra in the optical and infrared (IR) ranges
(NGC 7212, Contini et al 2012,  NGC 3393, Contini, 2012a, NGC 6240, Contini 2012b),
with the aim  of explaining the physical and chemical conditions of  gas and dust, and 
 looking for some records of the parent galaxy collision. We could find that collision records  appear
throughout the  extended narrow line region (ENLR) as isolated patches of matter in  abnormal conditions, 
such as high velocities, high  densities and relatively high ionization-level lines.

An important question about mergers  is whether  starbursts or AGNs are  the radiation  sources
(e.g. Gonzales-Alfonso et al 2004).
We have analysed  the spectra observed from the merger galaxies by models which account for
the  photoionizing flux from the AGN and from the starburst coupled with a shock wave hydrodynamic regime,
which is suitable to  the collisional  characteristic of  merging.
We have found that generally, the optical line spectra are  better reproduced  by the AGN,
while the IR line ratios are explained by the starburst.

The modelling of the spectra showed in particular that  very young stars, such as Wolf-Rayet, are present in the
centre of NGC 3393  whereas strong absorption from the Galaxy prevents to see them throughout the spectral energy distribution
(SED) of the continuum. The relative abundances of the heavy elements are an important issue in mergers because
they reveal whether primordial matter was trapped  at the time of collision.
We have found lower than solar metallicities in NGC 7212 and in NGC 3393 (except the high N/H
which derives from the Wolf-Rayet wind). 
In NGC 6240, the C, N, O and Ar abundances relative to H
appear higher than solar by a factor $\leq 2$, because H is trapped into  H$_2$ molecules
and therefore depleted from the gaseous phase.
On the other hand S/H, Si/H and Fe/H show lower than solar abundances characteristic of the ISM included in
the NLR  during the merging process. Actually the high shock velocities exclude the hypothesis of trapping
into grains due to sputtering.

In the present paper we will investigate Arp 220 (z=0.018), which
 is an ultraluminous infrared galaxy (ULIRG) with \LIR $\geq$ 10$^{12}$ \Lsol, at a
 distance of 72 Mpc (Graham et al 1990).
The merger  can be seen in  Graham et al (1990) observations. The image at 2.2 \mum shows two
resolved sources separated by a distance of 350 pc. 
The merging of two galaxies was revealed also by  the disturbed morphology and the tidal tails.
Graham et al discuss the distance of the nuclei and the age of the merger considering the
starburst, the AGN, dust and gas.
Taniguchi et al (2012)  on the basis of the two tidal tails, suggest a multiple merging system 
for Arp 220, composed of four or more galaxies arising from a compact group of galaxies.
The two compact radio peaks associated with the IR nuclei reveal  that they are radio sources and not
radio jets. 
 
Molecular gas and massive young stars might contribute by  $\sim$ 10\%  to the luminosity
(Sturm et al. 1996) of Arp 220. A hidden AGN similar to a LINER (Rieke et al 1985) is  revealed
by the broad Br$\alpha$ line (FWHM $\sim$ 1300 \kms, DePoy et al. 1987).
Rangwala et al. (2011) on the basis of observations by $Herschel$ claim that the large 
column densities observed for molecular ions strongly favour the existence of an X-ray 
luminous AGN (10$^{44} erg$ s$^{-1}$).

 Wilson et al (2006) found more than 206 centrally concentrated clusters which can be classified into
two groups : young and intermediate -age population of 10 and 300 Myr, respectively.
Armus et al (1989)  analysing the optical line fluxes for a sample of powerful far-infrared galaxies
including Arp 220, claim that a young stellar population is present in all of them. 

By modelling the spectra of Arp 220, we would like
 to add some information about mergers in general and Arp 220 in particular.
We will  carry on the investigation of the physical conditions and  the  abundances
of the heavy elements  throughout  the NLR of close Seyfert  galaxies which  
show merging, as was done for  e.g. NGC 7212 and  for NGC 3393 and  NGC 6240 that show a double nucleus.
Most observations of Arp 220 report on  molecular bands, we will focus instead on the lines emitted and absorbed by the gas
consistently with the continua emitted  and absorbed by gas and dust.
We will adopt  models which account for
photoionization by the AGN or by the starburst coupled to the shock hydrodynamics, which were
  assembled to  investigate the NLR of AGN and the starburst galaxy environments (e.g. Viegas \& Contini 1994
and references therein).

The nucleus of Arp 220  is heavily obscured (Smith et al 1989). The  depth of the 10 \mum
silicate band is only a lower limit. The observations of the continuum in the submillimeter  have shown
that the emission is optically
thick even at wavelenths longer than 100 \mum (Emerson et al 1984). This demonstrates that  extinction
can reach  Av=1000 mag. The mid-IR spectrum of Arp 220
indicates an active nucleus or plural active nuclei heavily obscured.
In agreement with the other mergers, Arp 220 has most probably a central starburst of  young stars.
We will check these prediction by modelling the spectra in the different frequency ranges.

The observed optical line spectra (Veilleux et al. 1999)
  are not rich in   number of lines,  in particular, the very significant  [OII]3727+3729 doublet is  lacking. 
So we will constrain the
models   recalling the spectra emitted by NGC 6240, 
another merger galaxy at low redshift with characteristics similar to Arp 220.
On the other hand, the continuum SED of Arp 220 is very well defined by  a rich collection of observational data.
The modelling of the line spectra must be consistent with the 
SED of the continuum, even if modelling the SED is less constraining (see Contini, 2012b).  
Previous models of  Arp 220 SED were presented by Contini et al (2004)
and by Contini \& Contini (2007)
on the basis of the data from the NED. The data were well reproduced in all
 the observed frequency ranges by a  composite model accounting for the flux from the AGN and shocks (Contini et al 2004)
and composite models of AGN and starburst and shocks (Contini \& Contini 2007).
Since then, the data from the  NED were enriched and updated.  Today, they  better delineate the main 
physical mechanisms at the basis of the Arp 220 SED, namely  bremsstrahlung from the gas, and reprocessed
radiation from dust. Yet, only the line spectra will decide whether  a power-law
radiation  from
the AGN or  the  black-body radiation from the starburst  characterises the  photoionization source.

The modelling process  is described in Sect. 2.
The comparison of the calculated line ratios with the data  appears in Sect. 3.
Dust and gas continuum SEDs  are presented and discussed in Sect. 4.
Concluding remarks  follow in Sect. 5.

\section{Modelling procedure} 

\subsection{About the observations}

In the optical range we will  refer to the spectrum  observed by Veilleux et al. (1999).
The spectroscopic data for Arp 220 are  presented within  the sample of IRAS 1 Jy sample of ULIRGs.
They were observed by the Gold Cam Spectrograph on the Kitt Peak 2.1m telescope.

In the IR we will model the spectrum reported by Sturm et al (1996)
 that  was obtained with the Short Wavelength Spectrometer (SWS) on board ISO.
 The data are constrained by the observations of 10-37 \mum spectra of ultraluminous infrared
galaxies (ULIRGs) by Farrah et al (2007) taken using the Infrared Spectrograph on board of {\it Spitzer}.
In particular, Sturm et al  do not detect high excitation fine
structure lines of [OIV], [NeV] and [NeVI] in Arp 220, suggesting  that such lines are emitted  from gas
photoionised by a much harder
radiation  than that emitted by the starburst, or heated by a relatively strong shock.
However their  conclusion derives from diagnostic diagrams, namely,
the  [NeII] line limits are below the ranges observed
by ISO in AGNs.
Sturm et al claim that although a low luminosity AGN could be present,
 star formation are the dominant source of radiation in Arp 220.
In fact  IR lines are generally emitted from gas photoionised by
 massive star formed in a recent starburst.

Rangwala et al. (2011) 
 detected two [CI] lines  and one [NII] line ([NII]205) in emission.
The ratio of [CI] line strength is 1.3 $\pm$ 0.3, which suggests that the lines are
optically thick, in agreement  with the excitation temperature of 26 K derived from
their temperature ratio (Stutzki et al 1997). [NII] 122 is observed in absorption.
They report L([NII])/L(FIR) $\sim$ 1.7 10$^{-5}$.  Malhotra et al. (1997) wondering about the low
  L[NII] relative to L(FIR)
 claim that [NII] and [CII] come  different regions, namely HII regions and   
photo-dissociation regions, respectively.
The [CII] 158 deficiency is generally found in ULIRGs.

Luhman et al (1998)
  presented measurements of the [CII] 157.74 \mum fine-structure line in a sample of ULIRGs
observed   with the Long Wavelength Spectrometer (LWS) on the Infrared Space Observatory (ISO).
They claim that the [C II] 158
line traces gas  ionized by stellar far-ultraviolet (far-UV) photons
with energies greater than 11.26 eV, the ionization potential
of neutral carbon. 
 The [CII] transition has a
critical density  of 3 10$^3$ and 50 \cm3 for collisions with
hydrogen and electrons, respectively (Flower \& Launay 1977;
Hayes \& Nussbaumer 1984).
Luhman et al. (2003)  reporting the  observations of the
 157.74\mum  $^2P_{3/2}$- $^2P_{1/2}$  fine-structure line of C$^+$ 
 claim that it is the
single brightest emission line in the spectrum of most galaxies,
providing as much as 1\% of the total far-infrared (FIR) luminosity
(see, e.g., Stacey et al. 1991 and references therein).
In Arp 220,  as in most  ULIRGs, the [CII] line  is unusually weak.
Only the [CII] 158 and OH 163 lines are clearly observed in emission. 
[OI] 63.2 is observed in absorption and 
[OI] 145.5 is not detected at all (Gonzalez-Alfonso et al. 2004).
%, and modest energy requirements

\subsection{The calculation code}

The line and  continuum spectra are calculated  by the code  
{\sc suma}\footnote{http://wise-obs.tau.ac.il/$\sim$marcel/suma/index.htm}
which simulates the physical conditions
of an emitting gaseous nebula under the coupled effect of photoionization from an
external source and shocks.
The calculations are  described in detail by  Viegas \& Contini (1994 and references therein) and Contini et al. 
(2012 and references therein).

The input parameters which characterise the model are : the  shock velocity \Vs, the atomic preshock density \n0,
the preshock magnetic field \B0. They define the hydrodynamical field and  are  used in the calculations
of the Rankine-Hugoniot equations  at the shock front and downstream. These equations  are combined into the
compression equation which is resolved  throughout each slab of the gas, in order to obtain the density 
profile downstream and consequently, the temperature.
We adopt  for all the models \B0=10$^{-4}$ gauss. In fact, a higher preshock density is compensated by a
higher magnetic field which reduces compression and viceversa. We do not have enough data throughout the Arp 220 ENLR
  to avoid degeneration  of models   choosing  between  \n0  and \B0 effects on the density (see Contini 2009).

The input parameter  that represents the radiation field is the power-law
flux  from the active center $F$  in number of photons cm$^{-2}$ s$^{-1}$ eV$^{-1}$ at the Lyman limit,
if the photoionization source is an active nucleus. The spectral indices are $\alpha_{UV}$=-1.5
and $\alpha_X$=-0.7.
 $F$  is combined with the ionization parameter $U$ by
$U$= ($F$/(n c ($\alpha$ -1)) (($E_H)^{-\alpha +1}$ - ($E_C)^{-\alpha +1}$)
(Contini \& Aldrovandi, 1983), where
$E_H$ is H ionization potential  and $E_C$ is the high energy cutoff,
n the density, $\alpha$ the spectral index, and c the speed of light.

If the radiation flux  is   black body radiation from the stars,
the input parameters are the colour  temperature of the  star \Ts
and  the ionization parameter $U$
(in  number of photons per  number  of electrons at the nebula).

The secondary diffuse radiation emitted from the slabs of gas heated by the shocks is also calculated.
The  flux  from the active centre,  from the stars and the secondary radiation are  calculated by 
radiation transfer throughout the slabs downstream.

The geometrical thickness of the emitting nebula $D$,
the dust-to-gas ratio $d/g$, and the  abundances of He, C, N, O, Ne, Mg, Si, S, A, and Fe relative to H
are also accounted for.
 The distribution of the grain radius  downstream
is determined by sputtering,  beginning with an initial  radius of  $\sim$ 0.5-2.5 \mum.

In our models the flux from an external source can reach the  shock front (the inflow case,
indicated by the parameter $str$=0) or the edge of the
cloud opposite to the shock front  when the cloud propagates outwards from the AC or from the starburst 
(outflow, indicated by $str$=1).

The fractional abundances of the ions are calculated resolving the ionization equations
considering ionization by the primary flux, the secondary flux and collisional ionization
in each slab downstream.
The  line intensity calculated in each slab  depends  strongly
on  the density,  temperature, and radiation from both the sides of the cloud. 
The line intensities are integrated throughout the downstream region up to a distance $D$  from
the shock front at which the gas has reached a relatively low temperature (T$<$ 10$^3$ K) in the case of
inflow (radiation bound case) or to a distance $D$ such that all the calculated line ratios reproduce
the observed ones (matter bound case).
In the outflow case some iterations are necessary to obtain converging results.

The models  which  reproduce the  observed strong lines within 20\% and the weak lines by 50 \%
are selected from a large grid of models (see Contini 2012b). The  relative sets of  imput parameters
are  regarded as {\it results}.  The approximation of calculated to observed line ratios
 should  account for the observational
errors and for the uncertainty of the physical coefficients adopted by the calculation code.

\subsection{Model selection}

\begin{table*}
\begin{center}
\caption{Comparison of calculated with observed optical spectra}
\begin{tabular}{lllllllll} \hline  \hline
   line     &obs$^1$                 &m1$_{pl}$&m2$_{pl}$ &m1$_{sb}$ & m2$_{sb}$   &m$_{sd}$   \\ \hline
\ \Hb       &  1.                & 1.      &1.        & 1.       &    1.       & 1.  \\
\ [OIII] 5007+ & 1.35            & 1.32     &1.34     & 0.9      & 1.1         & 3.     \\
\ [OI] 6300+  & 0.63             & 0.54    &0.81      & 0.68     & 0.13        &  0.25    \\
\ [NII] 6584+ & 7.2              &  7.14   &8.2       & 5.5      &  2.8        &0.5         \\
\ [SII] 6716+  & 2.93            &  2.9    &3.34      & 0.11     &  1.         &0.03     \\
\ \Hb (\erg)  &3.3e-15            & 7.5e-4  &7.3e-4    &0.21      &  0.0024     &0.039         \\ \hline
\  \Vs(\kms)  & -                 & 300    &400       &300        &400         &1000   \\
\  \n0 (\cm3) & -                 & 30     &30        &280        &40          &1000  \\
\  $F$ $^2$   & -                 &2.5e8   &2.5e8     &-          & -          &-  \\
\   $U$       & -                 &   -    & -        &20.        &8.e-3       &-  \\
\  \Ts (K)    & -                 & -      & -        &4.e4      &  4.5e4     &-  \\
\  $D$ (cm)& -                    &1.5e17  &1.5e17    &8.e18     & 3.e18       &1.e17     \\
\ N/H       & -                   & 1.4e-4 &1.5e-4    &1.5e-4    &1.5e-4       &9.1e-5 \\
\ O/H       & -                   & 3.6e-4 &3.6e-4    &9.5e-4    & 9.6e-4      &6.6e-4 \\
\ Si/H      & -                   & 2.0e-6 &2.0e-6    &3.3e-6    & 3.3e-6      &3.3e-5  \\ 
\ S/H       & -                   & 1.1e-5 &1.2e-5    &1.6e-6    & 1.e-5       &8.e-6  \\ \hline
\end{tabular}
\end{center}

$^1$ Veilleux et al (1998); $^2$ in photons cm$^{-2}$ s$^{-1}$ eV$^{-1}$ at the Lyman limit 

\caption{The IR calculated line ratios to [NeII] 12.8}
\begin{center}
\begin{tabular}{lllllllll} \hline  \hline
   line         &  obs$^1$ & obs$^2$ & m1$_{pl}$ & m2$_{pl}$ & m1$_{sb}$&m2$_{sb}$&m$_{sd}$   \\ \hline
\ [MgVII] 8.87  & -        & -       & 0.061     & 0.068     &   0.004  & 0.012   &  0.001  \\
\ [ArIII] 8.99  & -        &-        &  0.734    &   0.634   &  0.317   & 0.131   &  0.053 \\
\ [FeVII] 9.5   & -        &-        & 0.022   &  0.025    & 0.001      & 0.004   & 0.000\\
\ [SIV]   10.5  & -        &$<$0.02  & 0.067   &  0.066    & 0.000      & 0.011   & 0.000\\
\ [NeIII] 10.78 & -        & -       & 0.000   &    0.000  &   0.000   &  0.000   &   0.000\\
\ [ClIV]  11.76 & -        &-        &  0.001  &   0.001   &  0.000     & 0.000    &  0.000\\
\ [SIII]  12.0  & -        &-        & 0.000   &  0.000    & 0.000      & 0.000     & 0.000\\
\ [NeII]  12.78 & 1.       &1.       &  1.000  &   1.000   &  1.000    &  1.000     &  1.000\\
\ [ArV]   13.07 & -        &-        & 0.019   &  0.023    & 0.001      & 0.003     & 0.000\\
\ [MgV]   13.53 & -        &-        & 0.003   &  0.003    & 0.000      & 0.000      & 0.000\\
\ [NeV]   14.3  & $<$0.045 &         & 0.072   &  0.082    & 0.004      & 0.012     & 0.001\\
\ [ClII]  14.34 & -        &$<$0.04  &  0.006  &   0.004   &  0.001     & 0.003     &  0.005\\
\ [NeIII] 15.6  & -        &0.12     &6.828    &   5.536   & 13.230     & 0.390      &  0.776\\
\ [SIII]  18.67 &  $<$0.42 &0.084    & 1.088  &   1.272   &  0.127      & 0.147     &  0.007\\
\ [FeVI]  19.56 & -        &-        & 0.044  &   0.048   &  0.002      & 0.007     &  0.001\\
\ [ClIV]  20.35 & -        &-        & 0.001  &   0.001   &  0.000     &  0.000      & 0.000\\
\ [ArIII] 21.8  & -        & -       &0.054  &   0.047   &  0.020     &   0.009     &  0.003\\
\ [FeIII] 22.9  & -        & -       &0.253  &   0.150   &  0.212     &   0.095     &  0.259\\
\ [NeV]   24.17 &-         & $<$0.2  & 0.099  &  0.113    & 0.005      &  0.016      & 0.002\\
\ [OIV]   25.9  & $<$0.054 & $<$0.32 &  0.299   &  0.342    & 0.025   &   0.116          & 0.006\\
\ [FeII]  26.0  & -        &-        & 3.133   &  2.826    & 4.740    &   1.055         & 0.434\\
\ [OIII]  32.5  & -        &-        & 0.000   &  0.000    & 0.000   &    0.000         & 0.000\\
\ [ClII]  33.4  & -        &-        & 0.000   &  0.000    & 0.000   &    0.000           & 0.000\\
\ [SIII]  33.5  & 0.1      &1.16??   &  1.598   &  1.887    & 0.152  &    0.171         & 0.006\\
\ [SiII]  34.8  & 0.73     &0.5      & 0.647   &  0.57     & 0.743   &    0.273         & 0.043\\
\ [NeIII] 36.1  & -        & 0.74??  &  0.598  &   0.484   &  0.923  &    0.033        &  0.052\\
\ [OIII]  51.67 & -        &-        &  0.648  &   0.320   &  0.505  &    0.325         &  0.296\\
\ [NeIII] 57.3  & -        &-        &  0.427  &   0.204   &  0.028  &    0.114        &  0.064\\
\ [OI]    63.0  & -        &-        & 1.148   &  1.446    & 9.387   &    0.367       & 0.177\\
\ [NII]   76.07 & -        &-        & 0.000   &  0.000    & 0.000   &    0.000         & 0.000\\
\ [OIII]  87.3  & -        &-        & 0.478   &  0.207    & 0.069   &    0.146         & 0.039\\
\ [NII]   121.5 & -        &-        & 0.643   &  0.466    & 0.063   &    0.125          & 0.010\\
\ [OI]    147.0 & -        &-        & 0.095   &  0.123    & 0.559   &    0.028        & 0.011\\
\ [CII]   158.  & -        &-        & 0.739   &  0.800    & 0.159   &    0.118     & 0.016\\
\ [NII]   205.  & -        &-        & 0.144   &  0.102    & 0.007   &    0.0179   & 0.001\\ 
\ [CII]158/[NII] 205 & 17. $^3$ & - & 5.33   &  8.        &22.8    &     6.59     &314.8         \\ \hline
\ [NeII] (\erg)12.8&2.2e-12 $^1$ &6.5e-13&2.3e-4   &2.9e-4      & 0.028   &   1.38e-3   & 0.018        \\
\ [OI] 63 (\erg)  &-6.35e-13 $^4$ &-& 2.5e-4  & 4.2e-4     & 0.26  &    5.e-4      &0.0066        \\
\ [OI] 145 (\erg) &$<$0.94e-13 $^4$&-& 2.e-5  &3.5e-5      & 0.016 &    3.8e-5     &1.9e-4        \\
\ [CII] 158 (\erg)& 9.99e-13 $^4$ &- & 1.6e-4 &2.3e-4      & 0.004 &    1.63e-4   & 8.5e-5        \\ \hline

\end{tabular}

\end{center}
 $^1$ Sturm et al (1996); $^2$ Farrah et al (2007) $^3$ Rangwala et at. (2011); $^4$ Luhman et al (2003)

\end{table*}

A large grid of models was run in order to avoid degeneracy. In other words
we control  the line ratios so that they cannot be reproduced  within the allowed limits 
by models different from those selected (see Contini et al 2012  and references therein
for a detailed description of modelling).
  
A  first hint about the shock velocity is  given by the FWHM of the line profiles and about the gas density
by the [SII] 6717/6730 line ratios. 
Veilleux et al (1999)  measured FWHM$\geq$ 300 \kms for the Arp220 line profiles.
Then,
the first  evaluations  are slightly changed in order to obtain the best fit of all the line ratios to the data :
within 20\% for the strong lines and within 50 \% for weak lines.

Our method of modelling starts as usually by reproducing the oxygen line ratios, because oxygen is present with lines from
two or more different ionization levels. [OIII]/[OI]
is adopted in the present modelling instead of the [OIII]/[OII] line ratios which  are generally more significant, but
the [OII] doublet is  not  given by  the observations.
The [OI] line strength depends strongly on $D$, the geometrical thickness of the emitting cloud.
In matter bound models the [OI]/\Hb line ratios are not determined only by the physical
conditions throughout the cloud, but also by the geometrical thickness of the cloud.

So far the physical conditions of the emitting gas are determined. The relative abundances to H of 
elements which appear through only one line in the spectrum are calculated by changing the relative solar value
until the observed line ratio is satisfactorily reproduced.

 In Table 1 (top panel) we compare  model results with the optical  spectrum observed  
by  Veilleux et al (1999). The spectral line intensities were reddening corrected adopting E(B-V)=1.05.
The correction has been done on the basis of the Balmer line ratios. In {\it normal} conditions of
the emitting gas (densities between 10$^3$ and 10$^5$ \cm3 and temperatures between 10$^3$ and 10$^6$ K)
 \Ha/\Hb $\sim$ 3 (Osterbrock 1989). 

 In Table 1 we refer  the line ratios  to \Hb=1 in order to avoid problems of distances etc. . 
The absolute fluxes of \Hb observed and calculated by the models are
given in the last row of the top panel. 
Absolute fluxes are observed at Earth but calculated at the nebula, therefore they differ by
a large factor which depends on the distance of the nebula from the  AC and from the distance of the
galaxy to Earth (see Sect. 5).

In the bottom  panel of Table 1, we present the input parameters adopted
by the models, followed by the relative abundances C/H, N/H, O/H, Si/H, S/H etc which show some
variance from the solar ones  (C/H=3.3 10$^{-4}$, N/H=9.1 10$^{-5}$, O/H=6.6 10$^{-4}$, Ne/H=10$^{-4}$,
Mg/H=2.6 10$^{-5}$,  Si/H = 3.3 10$^{-5}$, S/H=1.6 10$^{-5}$, Fe/H=3.2 10$^{-5}$ Allen 1976).

The models were selected from a large grid of outflowing clouds in both the cases of an AGN or a starburst.
We tried to   obtain an approximation to the
observed line ratios within 50 \%. We chosed the models in order to show how  single parameters
affect the results. The power-law flux from the AGN is  considered for  models m1$_{pl}$ and m2$_{pl}$.
The black body flux from the starburst is the photoionising source for models m1$_{sb}$ and
m2$_{sb}$. 
 Recall that the models account for the coupled effects
of an external photoionization source
 and  shocks  (Sect. 2.2).  In the last column of Table 1 we present the results of a shock dominated model
m$_{sd}$ (F=0, U=0) calculated with \Vs=1000 \kms and \n0=1000 \cm3 which were found adapted to
reproduce the IR bump of the continuum SED (Contini et al. 2004).

 In Table 2  we report the  spectra
presented by Sturm et al (1996) corrected for Av=50 mag and those presented  by Farrah et al (2007) which were not corrected.
Farrah et al (2007) in their table 2 show that the correction factors for line ratios within the 12.8-24.3 \mum range
 for different extinction laws, are  within the observational error.
In the FIR range   extinction  drops to  zero (Draine 2009, fig. 1).
The FIR  data  come from Luhman et al (2003) observations 
of [OI] and [CII] from LWS ISO
and the  [CI] lines at 492 and 809 GHz and [NII] at 1462 GHz were  observed by Rangwala et al (2011).

 The models which appear  in Table 1 were selected  in order to explain  the optical lines, therefore they 
refer to \Hb, whereas
in the IR and far-IR ranges the lines are referred to [NeII] 12.8 which is
a relatively  strong line (Table 2). Moreover, Ne appears through lines in different ionization levels.
Most of the observed IR line ratios are upper limits. 
We will show in the following that the [NeIII]15.6/[NeII]12.8 line ratio
can be used to constrain the  models.

In Table 2  the  IR line ratios  are calculated by the same models
as those presented in Table 1,
 but the optical and infrared modelling are not normalized together
  because   the observational  spectra do not  provide  the data from
optical to far-IR consistently. 
 For a complex ensemble of different physical and chemical conditions such as the Arp 220 galaxy,
different  observations from different locations would be useful. We deal here with summed spectra
while  it is clear that  optical and IR lines could come from different places.
The low E(B-V) deduced by Veilleux et al shows that there are clouds reached by the AGN radiation,
 less obscured which can be seen in the optical range.
In fact  (Table 1)  dusty clouds emitting the IR spectra
have geometrical thickness not exceeding 1 pc, corresponding to fragmented  matter at the shock front.

  In the  last line of the top panel of Table 2 
we refer to  [CII]/[NII].  In the  bottom  panel of Table 2
the absolute fluxes provided by the observations of some significant IR lines are reported as well as 
model results.

\begin{figure*}
\includegraphics[width=8.8cm]{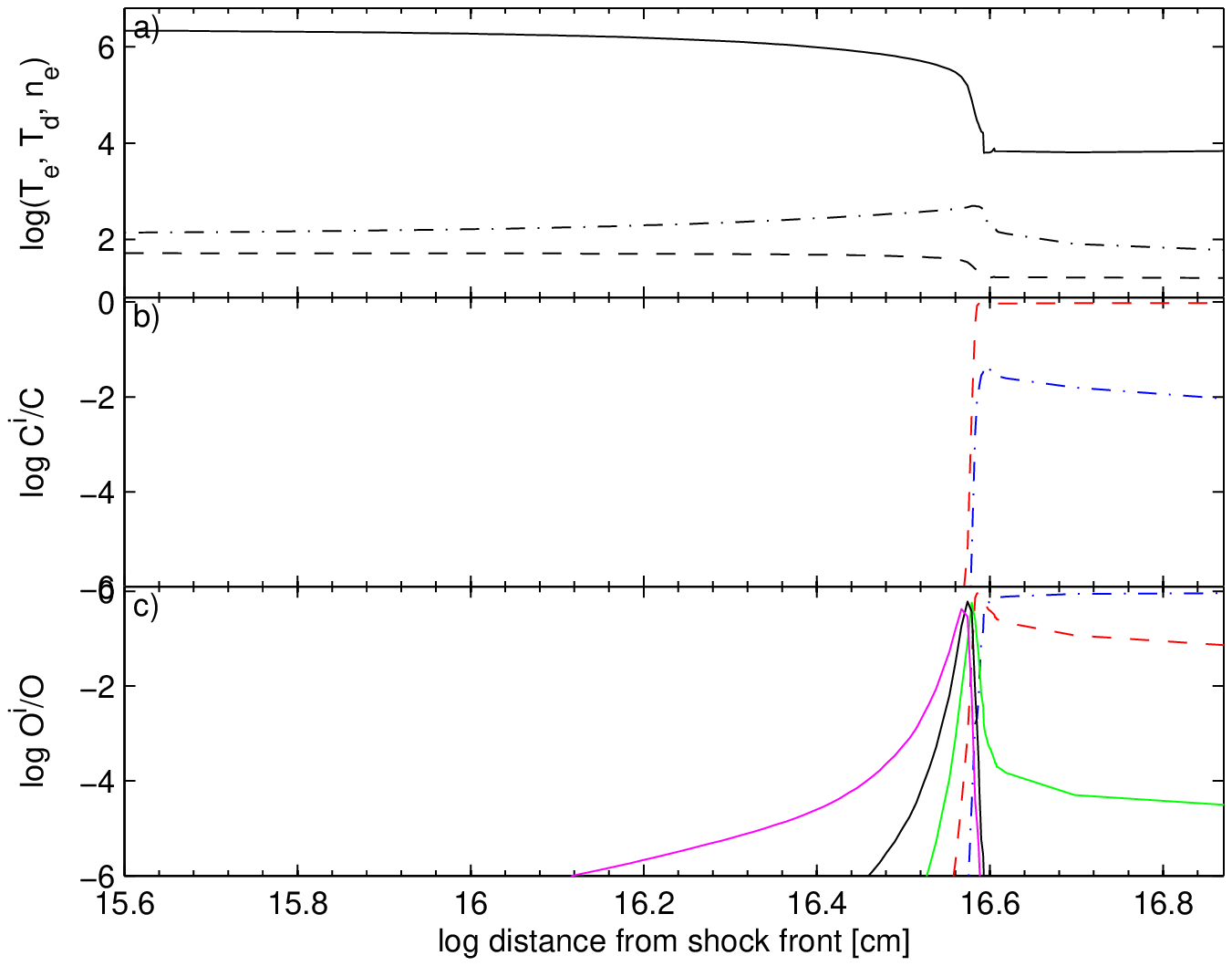}
\includegraphics[width=8.8cm]{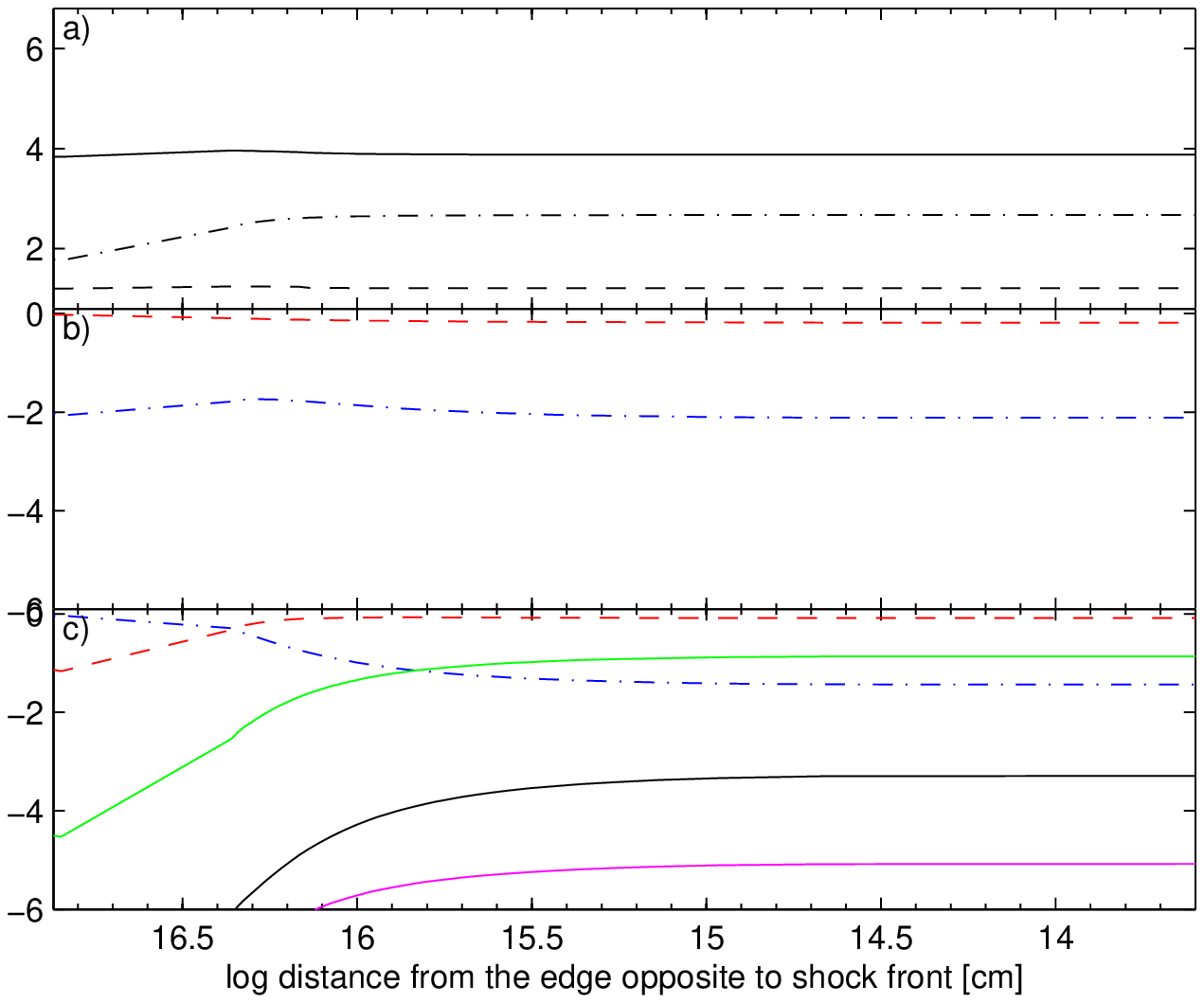}
\caption{
The physical conditions throughout a cloud corresponding
to model m1$_{pl}$ (Table 1). The shock front is on the left of the left
diagram; the
cloud edge reached by the flux from the AGN is on the right of the right
diagram (see text).
Top panel : \Te (solid line) , \Td (dashed line), \ne (dash-dotted line) .
Middle panel :  The fractional abundance of the  carbon ions : blue dash-dotted:
C$^0$/C; red dashed : C$^+$/C;
Bottom panel : blue dash-dotted: O$^0$/O; red dashed : O$^+$/O;
green : O$^{++}$/O; black : O$^{3+}$/O; magenta : O$^{4+}$/O.
}
\includegraphics[width=8.8cm]{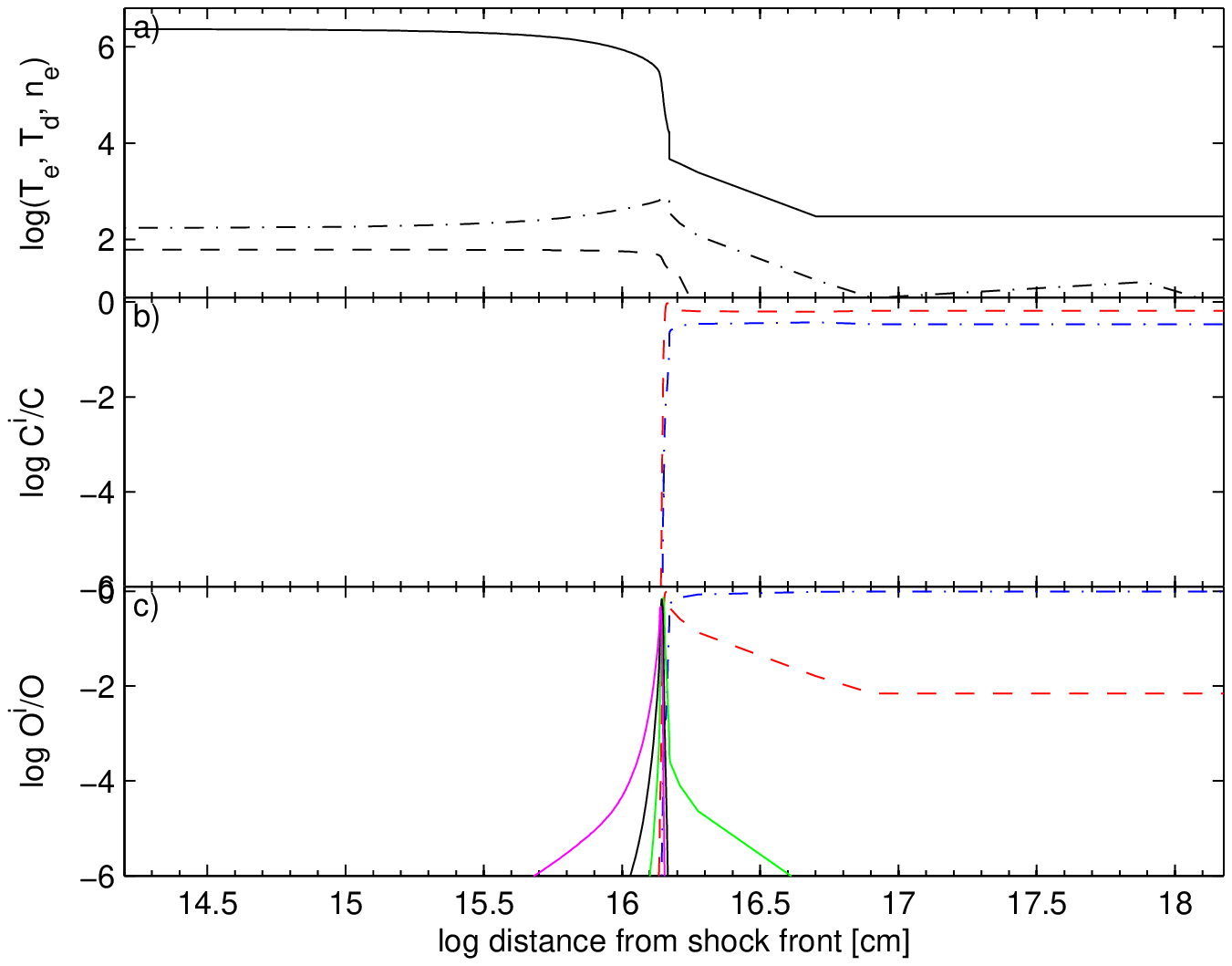}
\includegraphics[width=8.8cm]{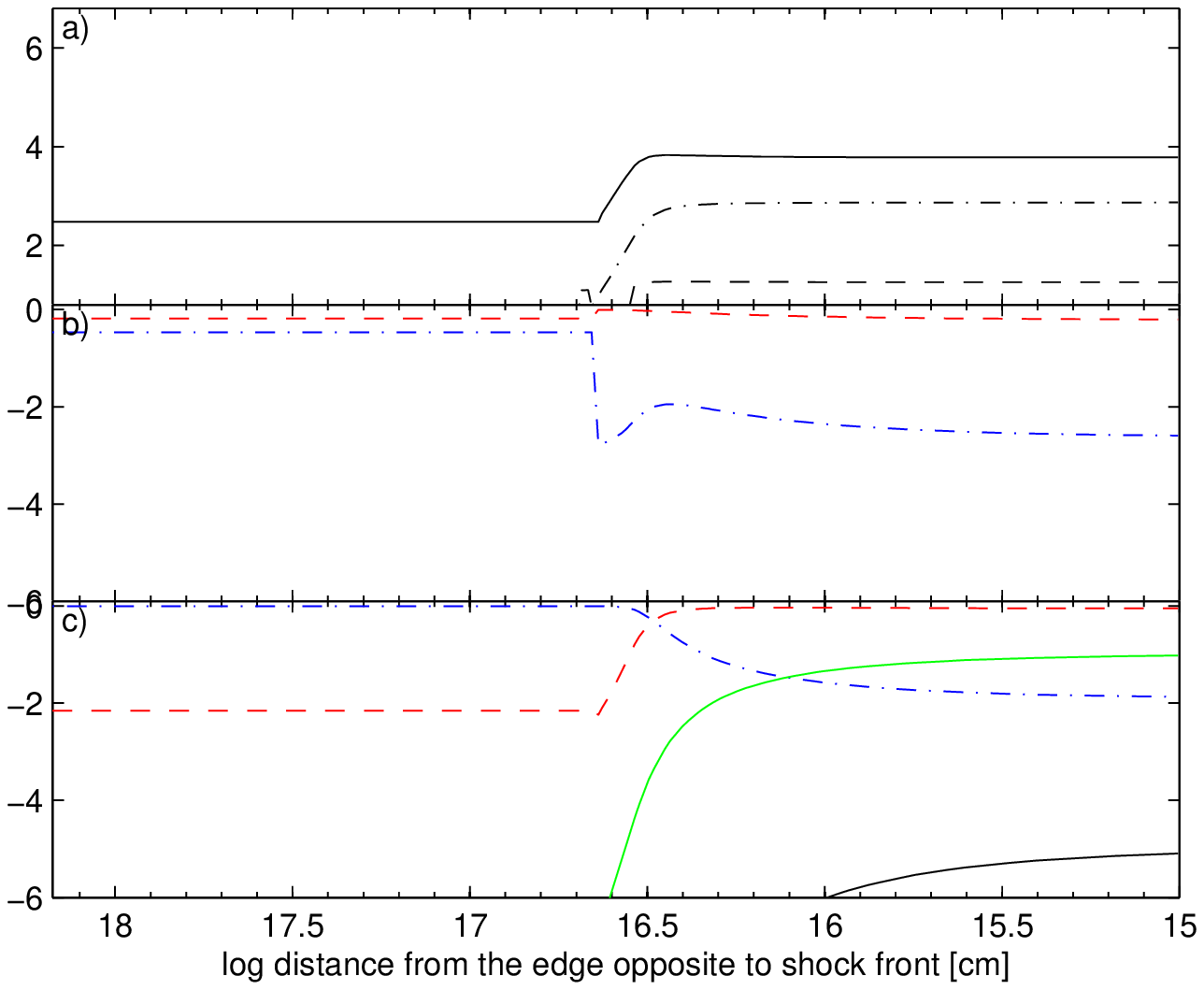}
\caption{
The same as for Fig. 1 for the m2$_{sb}$ model.
}
\end{figure*}

\section{The line spectra}

Examining the data of the optical spectrum  reported in Table 1,  Veilleux et al. (1999) 
 could not decide  whether the energy source  in Arp220 is accretion into
a massive black hole, or a starburst, or even shocks.
We have demonstrated in previous works (e.g. Contini et al. 2002a and references therein) that shocks 
accompany the outward motion of clouds in the NLR of Seyfert galaxies, LINERs etc., even if radiation
from the AC is the main photoionization source. In starburst galaxies the shock hydrodynamics is coupled to
radiation from the stars. In merger Seyfert galaxies shocks are created by   galaxy interaction,
double AGN were observed and starbursts arise from collision  of  galactic  matter.
We suggest that all of the three energy sources should be accounted for  in modelling the  Arp 220  spectra.

\subsection{Optical lines}
Calculation results show that the models which reproduce  within the  accepted approximation 
the optical lines (Table 1) account for a shock velocity of 300-400 \kms. The shocks are coupled  to a 
photoionization  power-law (pl) radiation flux corresponding to  
the  AGN, in particular  a LINER. In fact the 
flux intensity is rather low ($F<10^9$ photon cm$^{-1}$ s$^{-1}$ eV$^{-1}$ at the Lyman limit, Contini 1997) yielding
 a relatively low [OIII]5007/\Hb.
The  low photoionizing radiation flux from the AGN  is similar to that found in NGC 6240 (Contini 2012b).

The models selected  to represent the gaseous clouds  reached by the AGN radiation flux
 (m1$_{pl}$ and m2$_{pl}$  in Table 1) refer to outflow. They reproduce the observed line ratios 
within 15 \%.
Model m2$_{pl}$ calculated with  \Vs=400 \kms shows that a higher \Vs   increases the  temperature
 of the emitting gas, enhancing the lines corresponding to higher ionization levels (e.g. [OIV] and [NV], Table 2).

Starburst models (m1$_{sb}$ and m2$_{sb}$ in Table 1) are calculated with shock velocities similar
to those adopted for pl dominated models, but   the photoionizing flux is 
a black body radiation  from the stars.
Starburst (sb) models are calculated  by a black body radiation 
corresponding to \Ts=4.0 - 4.5  10$^4$ K.
The temperature of the stars is about in the norm (1-5 10$^4$ K).
The clouds are outflowing.
The corresponding spectra  generally  show  weak low-level and neutral lines, due to the exponential
character of the radiation flux.

 From the
geometrical thickness of the m1$_{pl}$ and m2$_{sb}$ clouds ($D$ $\leq$ 1 pc), we understand that  matter 
is strongly fragmented by the underlying turbulence at the shock front.

The optical line ratios corresponding to the sb models shown in Table 1 hardly  fit the  data.
 Model m1$_{sb}$ is  anomalous, because to
reproduce the relatively high [OI]/\Hb  an unusually high ionization parameter U=20 has been adopted.
Even so,  to reproduce the observed  [SII] lines  we would need  S/H  more than twice solar.
This is unacceptable in general and in particular for Arp 220
where a large extinction has been measured  and the observed 9.7 \mum absorption feature
indicates that S  may be locked up in silicate grains (Draine 2009).
We have  constrained the S/H relative abundance  by the IR line ratios.

Cross-cecking the IR spectra,
model m1$_{sb}$  overpredicts the [NeIII]15.6/[NeII]12.8 line ratio by
a factor $\geq$ 100.  
The only way to  lower this  line ratio is to reduce drastically
\Ts and/or U, which, however, yields a strong decrease of [OI]/\Hb. Considering the discrepancies
between calculated and observed data due to observational errors,   approximation of
the atomic parameters  used by the models etc., we   choose  model m2$_{sb}$
as representative of the starburst models.
Model m2$_{sb}$ calculated by U=0.008  shows discrepancies $>$ 60 \% between  calculated and observed
low-ionization level and neutral line ratios. For [OIII]/\Hb  the discrepacy is 19 \%.
This model   is a compromise between
unacceptably low line ratios in the optical range and unacceptably high [NeIII]15.6/[[NeII]12.8 in the IR.

Tables 1 and 2 show that model m1$_{pl}$ better explains the optical spectrum, whereas model m2$_{sb}$
satisfactorily reproduces the IR line ratios. However, the nebulae  emit both  optical and IR lines.
We will show in the following that the optical spectrum from  m2$_{sb}$
and the IR spectrum from  m1${pl}$ (which  also strongly overpredicts  [NeIII] 15.6/[NeII] 12.8) 
do not contribute to the final total   spectra.

 Applying  to the m2$_{sb}$ optical spectrum an extinction corresponding to E(B-V) $\sim$50
 (Sturm et al. 1996) in the heavily obscured starburst region,
the optical [OIII]/[OI] line ratio will sharply  decrease (Draine, 2009, fig. 1), but more interestingly,
the intensity of all the single lines will drop to zero considering 
that   I$_{\lambda}$=I$_{\lambda 0}$ e$^{- {\tau}_{\lambda}}$
where ${\tau}_{\lambda}$ = C f($\lambda$) (C=E(B-V)/0.77 and the extinction function f($\lambda$) in the optical range 
between \Ha and \Hb is of the order of 1 (Osterbrock 1989)). 
The optical lines from the  starburst region  will be unobservable.
We can therefore accept a less accurate fit  of the optical spectrum by models referring to the starburst
and, reducing U $<$  0.008,  reproduce the [NeIII]/[NeII] line ratio within a factor of 20 \%.
Concluding, the optical spectrum corresponds to the AGN, even if the AGN is actually hidden.
This will be discussed in Sect. 5.

\subsection{IR lines}

The IR  observations of Arp 220 presented by Sturm et al (1996) and Farrah et al (2007)
  are adopted  to model the IR lines. They appear in the second and third column of Table 2,
respectively.
To avoid problems of relative abundances, distances etc. we   refer the IR line ratios
to [NeII] 12.8. A large number of lines from different elements and different ionization levels
were calculated and are presented in Table 2. 
The observed IR lines are hardly  affected by extinction whichever E(B-V),
because for $\lambda$ $\geq$ 10 \mum  f($\lambda$) tends to 0. 
The models are the same as those adopted for the
calculation of the optical spectra (Table 1 bottom).
 The models referring to the starburst
were constrained mostly by the upper limits of the IR data, while the models referring to the AGN
were constrained by the observed optical line ratios.
The IR line ratios are satisfactorily reproduced by starburst models corresponding to a black body  flux
corresponding to \Ts=4.0 - 4.5  10$^4$ K.

Notice that the absolute flux of [NeII] calculated by  model m1$_{pl}$ is lower by a factor of $\geq$ 6 
 than that  of m2$_{sb}$. Moreover, summing up the  IR  lines of the two models m1$_{pl}$ and m2$_{sb}$,
the contribution of the IR spectrum from the starburst region will dominate adopting 
a  weight $W$=1 for the starburst contribution and  $W$=0.01 for that of the AGN. 
Also the shock dominated model m$_{sd}$ shows [NeIII]15.6/[NeII]12.8 higher than
observed by a factor of $\sim$ 6.4, 
therefore it should be added to the other contributions by a  weight $\leq$ 0.1.
This results agree with the starburst characteristic of the Arp 220 galaxy.

The modelling of the IR spectra also shows that 
   Fe/H  should be depleted from the gaseous phase by a factor $>$ 10.
The O/H relative abundance is about half solar in the clouds reached by the AGN flux,
while in  the starburst environment  it  is higher than solar by a factor of 1.5
relatively to the Allen (1976) values but roughly solar adopting  Anders \& Grevesse (1989) O/H=8.5 10$^{-4}$.
N/H is  higher than solar by a factor of $\sim$ 1.5 for both  AGN and starburst clouds.
Si/H is strongly depleted because locked into silicate grains. 
S/H is slightly depleted in clouds reached by the AGN flux, but it is reduced by a factor of 10
in the starburst clouds consistently with  the strong obscuration.
We suggest that the optical lines are emitted from clouds reached by the power-law flux from the AGN, even
if the active nucleus (or active nuclei) is (are) hidden by a large amount of dust

\begin{figure*}
\includegraphics[width=7.0cm]{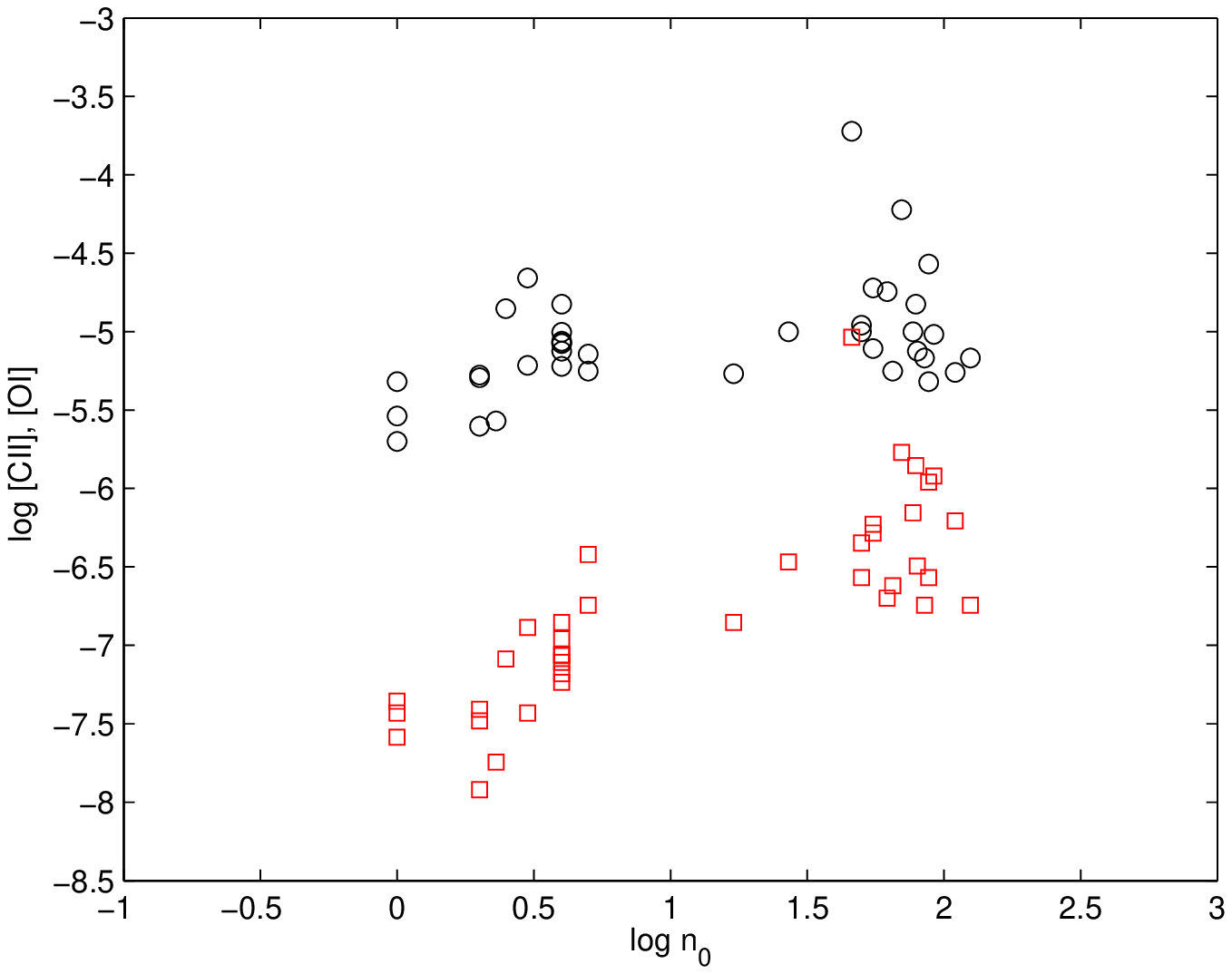}
\includegraphics[width=7.0cm]{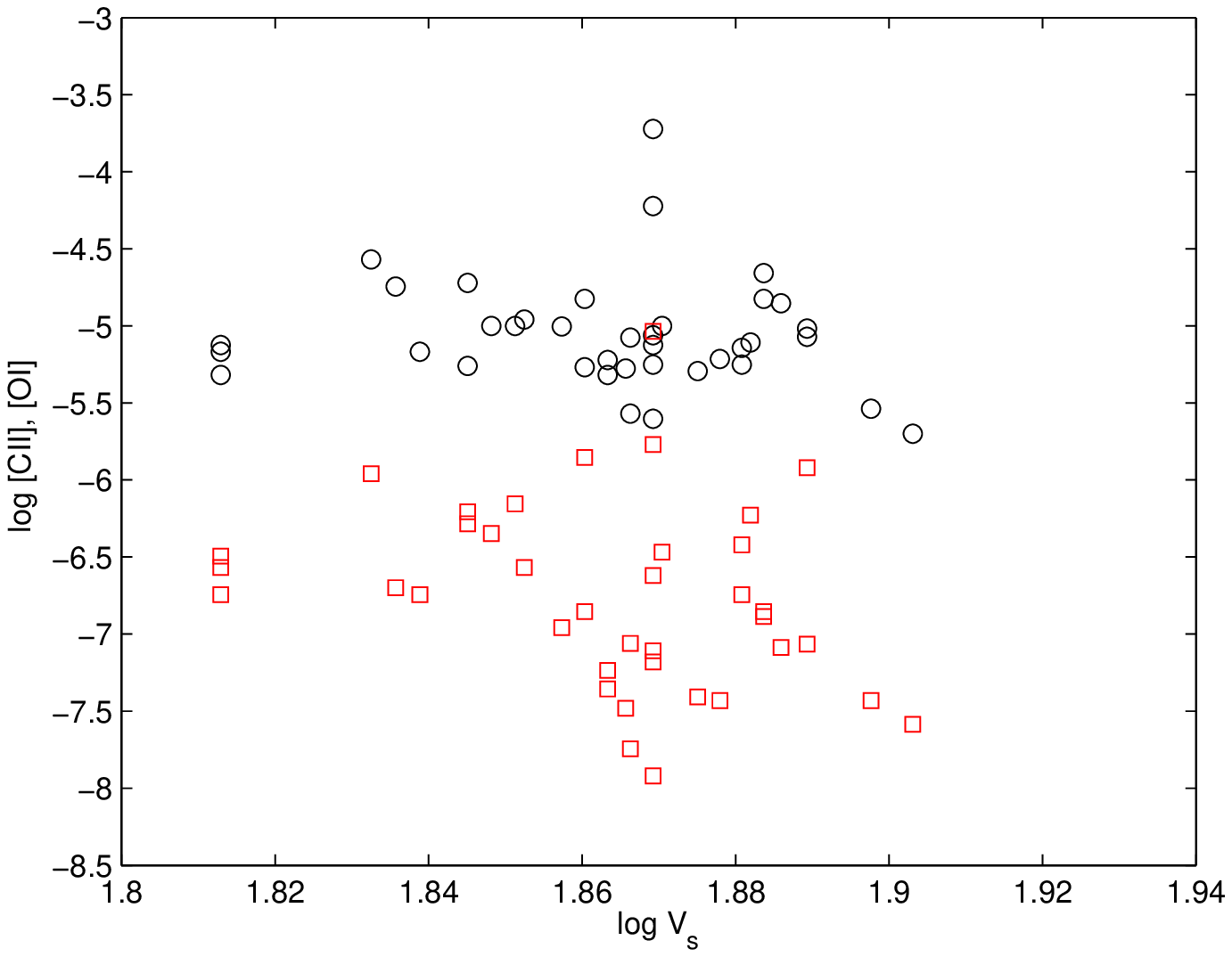}
\includegraphics[width=7.0cm]{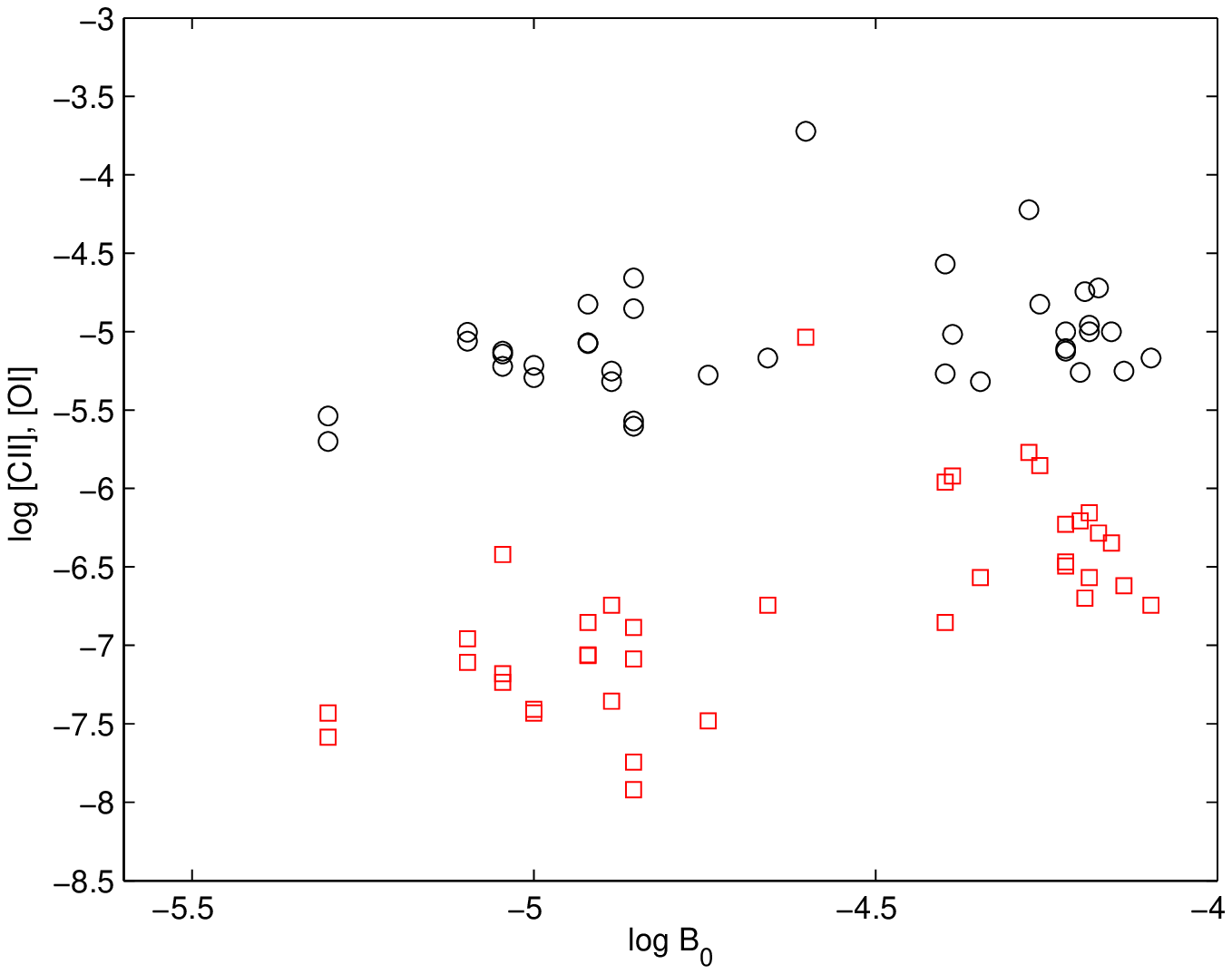}
\includegraphics[width=7.0cm]{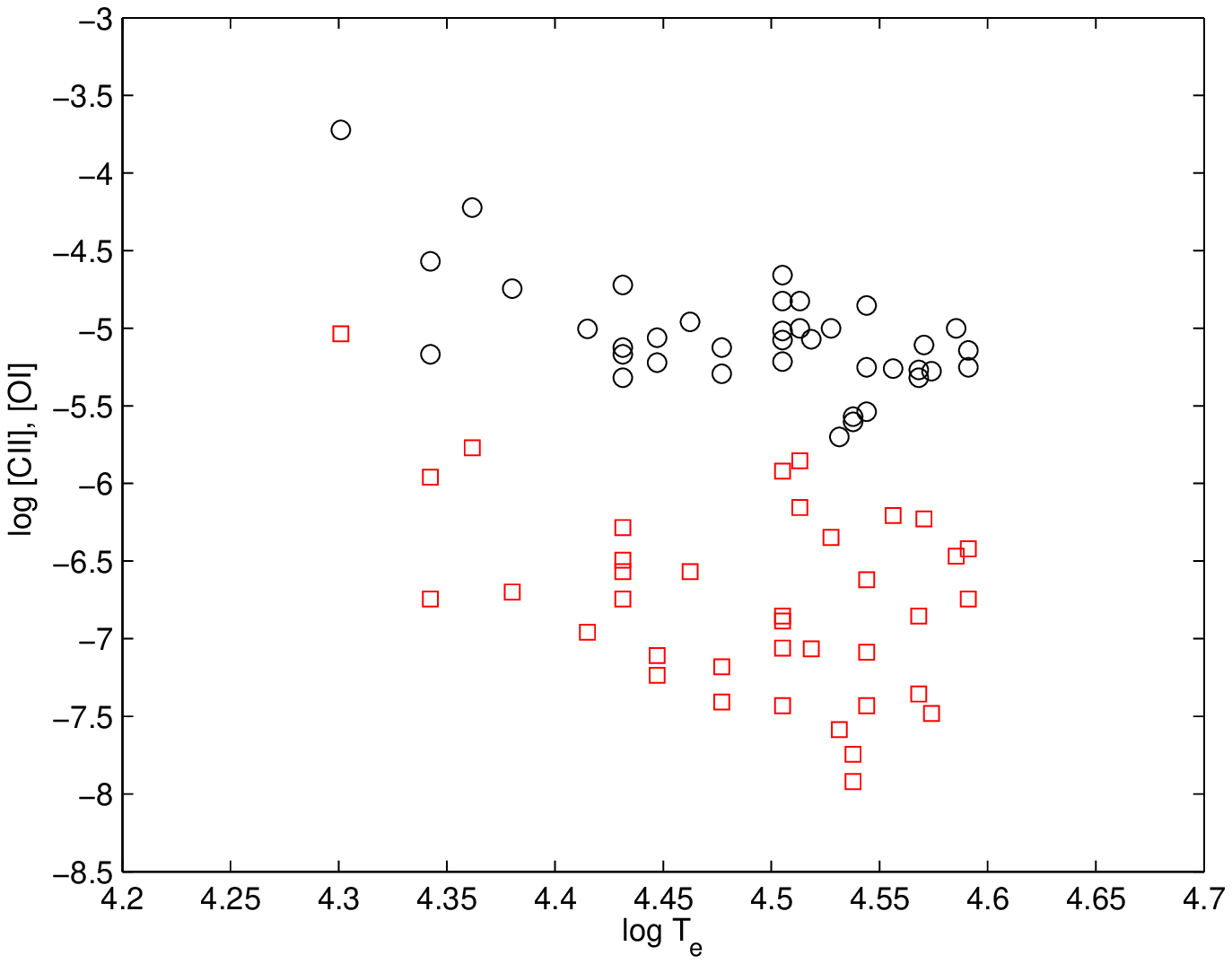}
\includegraphics[width=7.0cm]{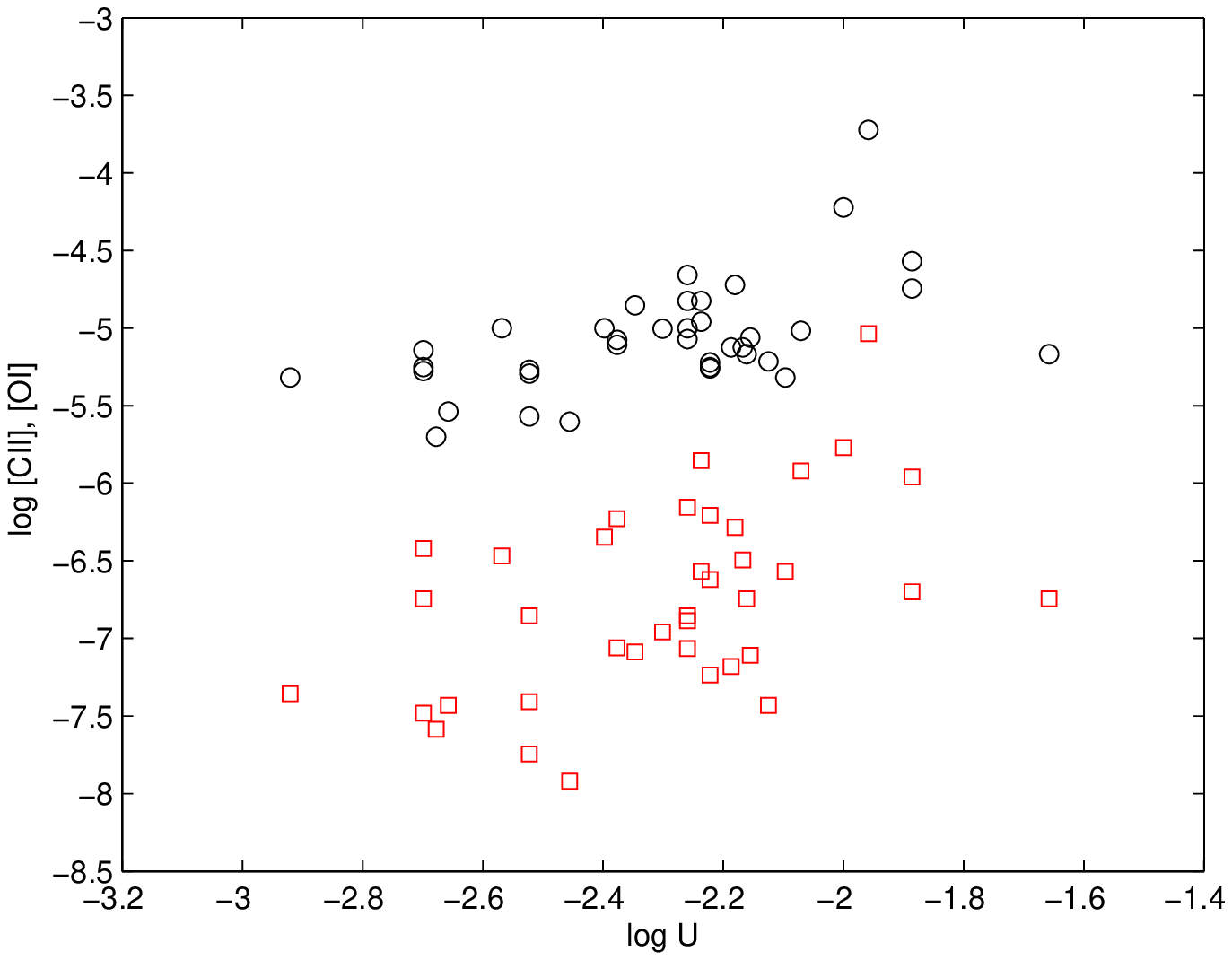}
\includegraphics[width=7.0cm]{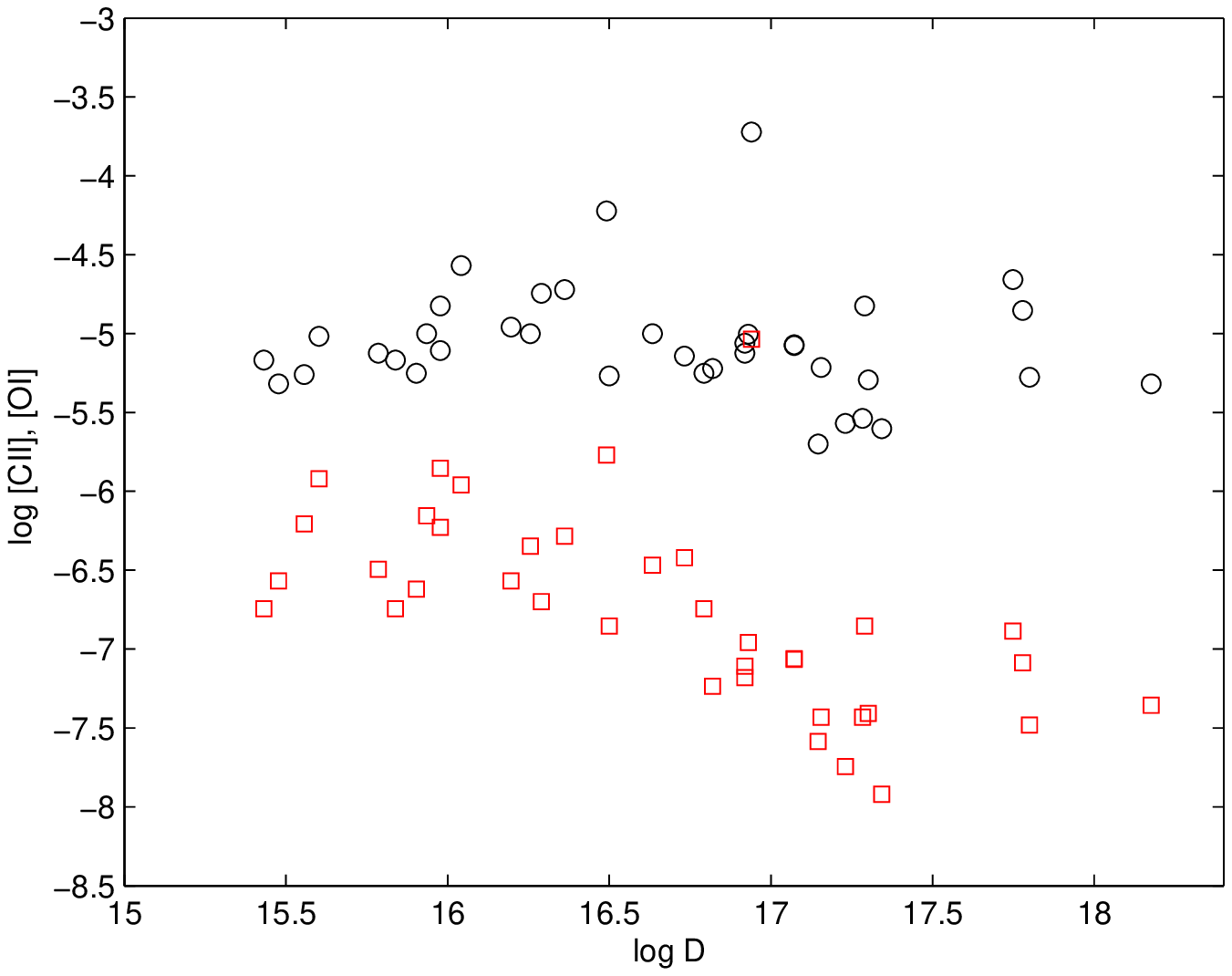}
\caption{The [CII]157 (black circles) and [OI]63 (red squares) lines (in \erg) calculated in 38 regions 
near the Galactic centre
(Contini \& Goldman 2011) as a function of different parameters.
}

\includegraphics[width=7.0cm]{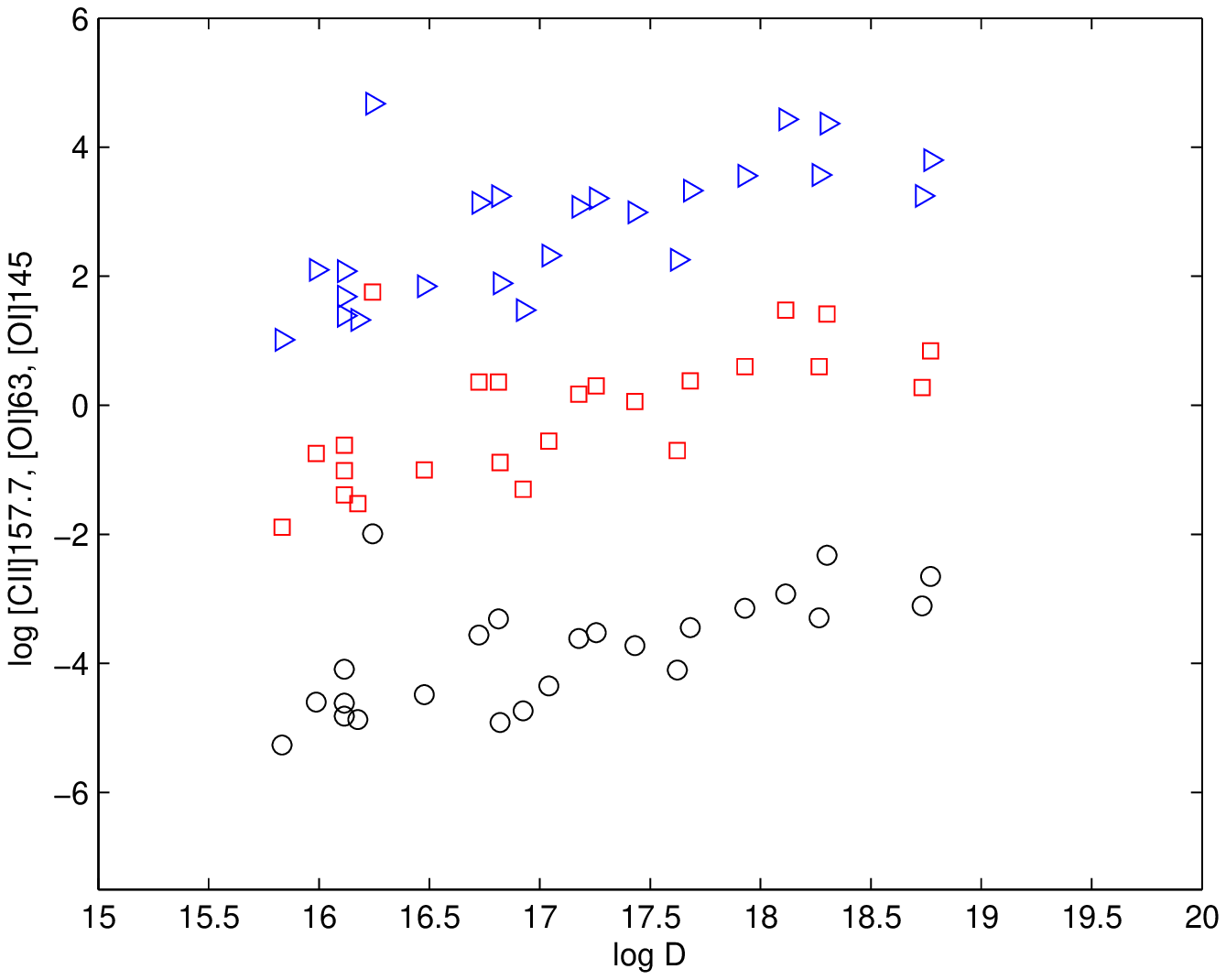}
\includegraphics[width=7.0cm]{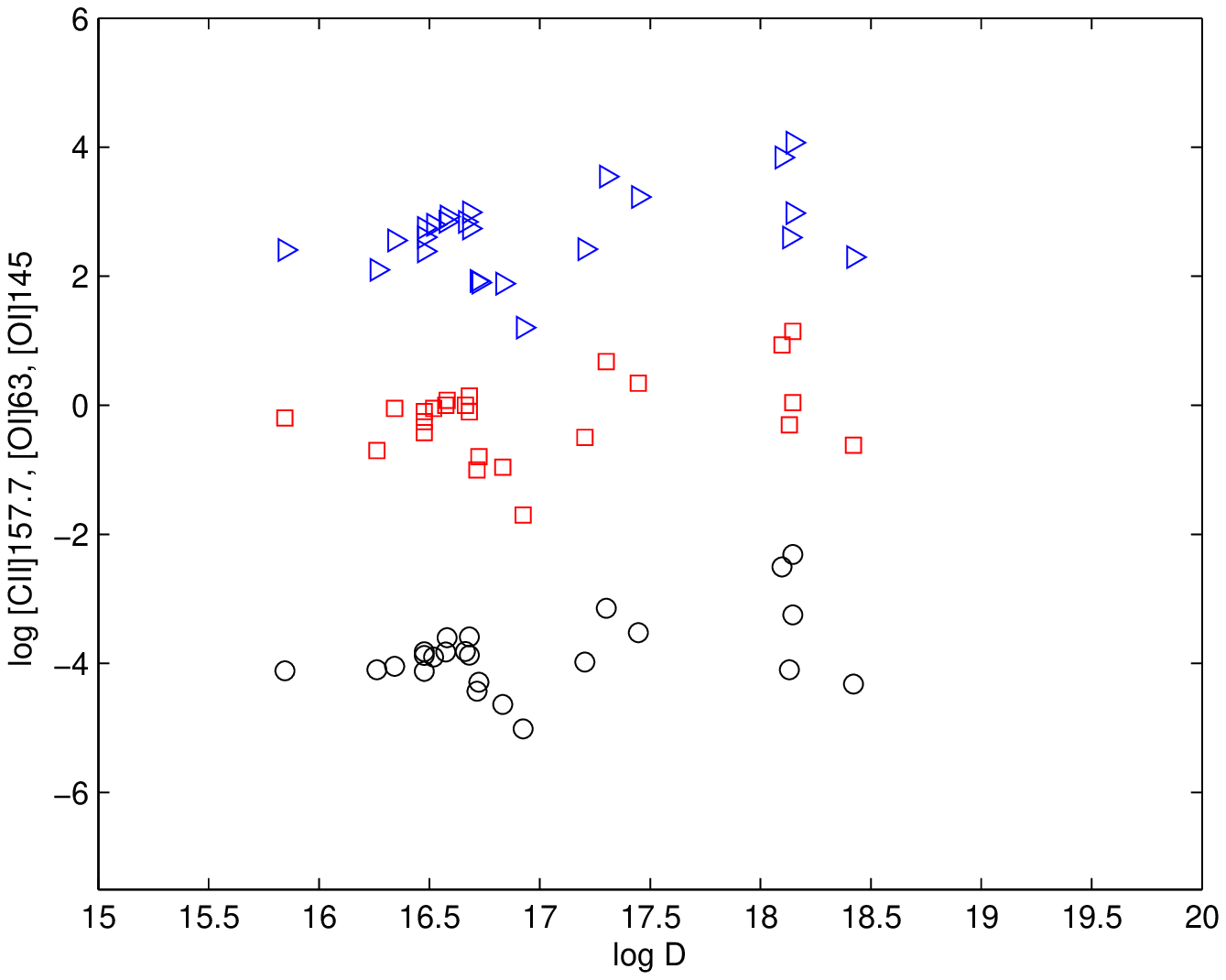}
\caption{The [CII]157 (black circles) and [OI]63 (red squares), [OI] 145 (blue triangles) lines (in \erg) calculated
in the Galactic centre by pl dominated models
(Contini 2011) as a function of $D$ (left panel). The same for bb dominate models (right panel)
}
\end{figure*}

\subsection{Far-IR lines}

In Table 2 (bottom panel) we  report the  absolute fluxes of the far-IR [OI]  and [CII] lines observed and
calculated  by the same  models which were used to model the optical and IR line ratios.
The [CII]/[NII] line ratios depend not only on the physical conditions of the emitting gas, 
but also on the relative abundances. 
Table 2 shows  that model m2$_{sb}$ underpredicts the observed  line ratio by a factor $>$ 2.
A N/H solar abundance would improve the fit.

In order to understand why some neutral lines are seen in absorption and low ionization lines are seen in
emission, we present in Fig. 1 the physical structure of the clouds corresponding 
to models m1$_{pl}$ and m2$_{sb}$.
In Fig. 1 we show the profile of the physical quantities , \Te, \Td ~ and \ne (electron temperature, dust temperature
and electron density, respectively)
and of the main fractional abundance of some significant ions throughout the clouds.
 Fig. 1  refers to the cloud reached by the pl flux from the AGN and  Fig. 2 to the cloud reached by
the black body (bb) ionizing flux  from the starburst.
Figs. 1 and 2 are divided into two  halves in order to show the critical zones
of recombination on both  the edges (the shock front is at the left of the left panel and the edge reached
by the photoionizing flux is at the right of the right panel). 

The temperature profile is different throughout the  pl and bb flux dominated  clouds. After the steep drop
due to gas recombination when the temperature reaches $\leq 10^5$ K, \Te remains $\sim$ 10$^4$ K due to secondary radiation
throughout most of the cloud in the pl case,  while  \Te $<$ 10$^3$ K throughout  most of the internal 
region of the bb radiation dominated cloud.
Figs. 1 and 2 show that  most of the neutral and low ionization level lines are emitted from the photoionized
side of the  cloud.  In the pl case the [CII] lines  are much stronger than  [CI].
In fact,   C$^0$/C  (Fig. 1, middle panel) is relatively low throughout all the pl dominated cloud.
Relatively strong  [OII]  can be predicted  in the spectrum  presented in Table 1 
   because also [NII]/\Hb is relatively high, not only due to a high N/H.
On the other hand,  [CI] and  [CII] are both strong in the internal region of the bb cloud (Fig. 2)
while [OI] is stronger than [OII].
In the bb  case, the clouds are  geometrically thick  ($D$=3. 10$^{18}$ cm) and optically thick
because the emission measure is high ($\Sigma$ n(i) \ne(i)  $\Delta_x$(i) $\sim$ 3 10$^{24}$ cm$^{-6}$ cm,
where $\Delta_x$(i)  is the single slab thickness),
therefore  [CI] and [OI] lines are both absorbed   throughout the internal region of the cloud. 
However, on the photoionized 
side of the bb cloud, at $\sim$ 3 10$^{16}$ cm from the edge, the [CI] and [OI] lines drop while the [CII] and
[OII] lines can be observed, although weak.

So far we have explained the [CI], [CII], [OI] line stength by the models which fit   Arp 220 spectra.
Now, extrapolating to Arp 220, we will consider  the FIR observations of another galaxy,  namely  the Milky Way
because many  spectra are available at close positions near the Galactic Centre (GC)
(Simpson et al 2007) and throughout the GC (Yasuda et al 2008). We investigate
 which  of the input parameters adopted  by the models which fit the observed IR spectra 
 could explain the  relatively weak [CII] 158 line  observed in emission and the
absorbed  [OI] 63 line.
We will use the results of the models calculated
in  previous works  (Contini, 2009, Contini \& Goldman 2011), where
we  presented the line ratios in the UV, optical and IR based on the models which explain 
the  IR observations  of the GC. The emitting clouds are   
photoionised by the bb flux from the star clusters and heated by the shocks. 

Although the physical conditions in the GC are not always similar to those found for Arp 220, we 
  report in Fig. 3 the trend of [CII] and [OI] lines as function of the single input parameters,
We would like to find out which parameter can  explain   the decrease of the [OI] lines relatively to [CII].

Fig. 3 shows that the preshock density combined with the  geometrical thickness of the
emitting clouds leads to a clear difference in the trends of [OI] and [CII],  diverging at  large $D$.
We find that both lines are reduced at high \Ts, low $U$, low \n0, therefore ULRIGs correspond
to relatively young starbursts, but the different trends  shown by the [CII] and [OI] lines
indicate that  the absorption of the [OI] line is due to a large geometrical thickness ($D$) of the clouds,
leading to a large optical depth. 
However, when the geometrical depth is very large ($\geq 10^{19}$ cm) and the flux is a pl from the AGN, 
the [CII] and [OI] lines both increase with $D$. 
This appears in Fig. 4 where the line intensities are calculated for the clouds  throughout the GC,
adopting an AGN as the source of photoionization (Contini 2011). The calculated lines are constrained by the
observations of the FIR lines by Yasuda et al (2008).

\begin{figure*}
\includegraphics[width=7.0cm]{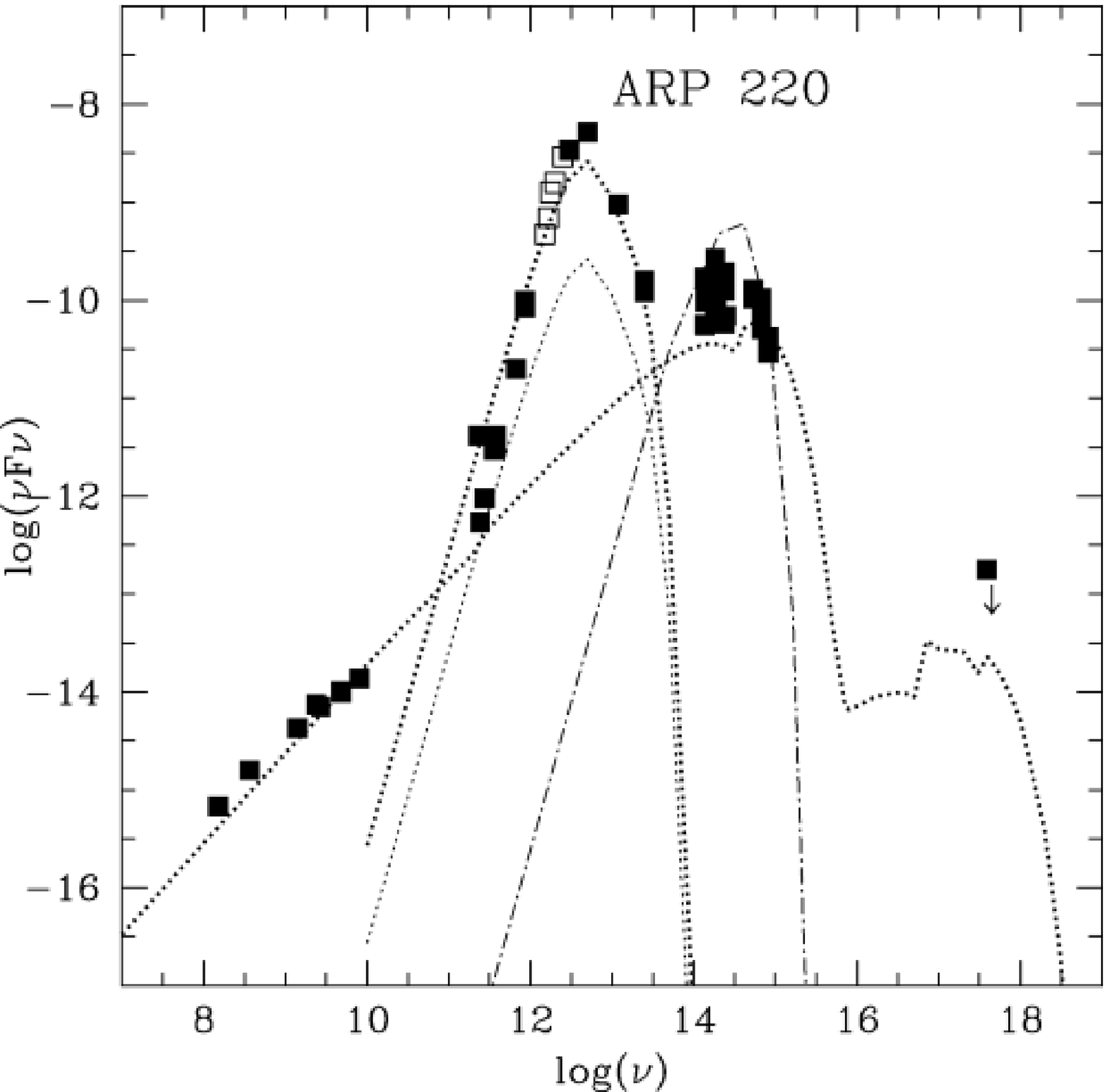}
\includegraphics[width=7.5cm]{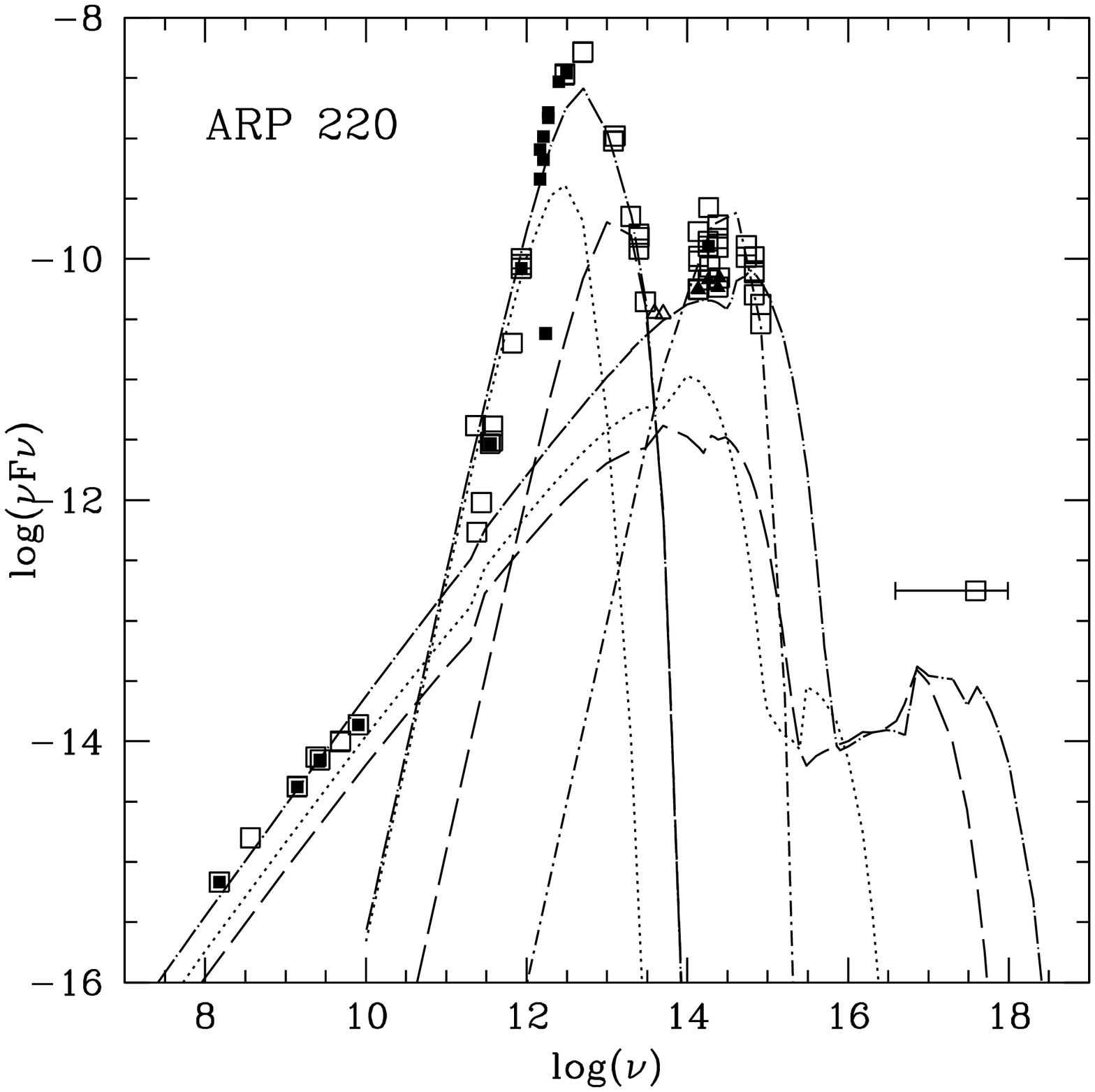}
\caption{Previous modelling of Arp.220. Left: Contini et al (2004, fig.A, symbols as in fig. 7, 
corresponding to table 1). Right :
Contini \& Contini (2007, fig. 6,
symbols are given in the caption of fig. 5 and refer to table 2)
}
\end{figure*}

\begin{figure*}
\includegraphics[width=8.cm]{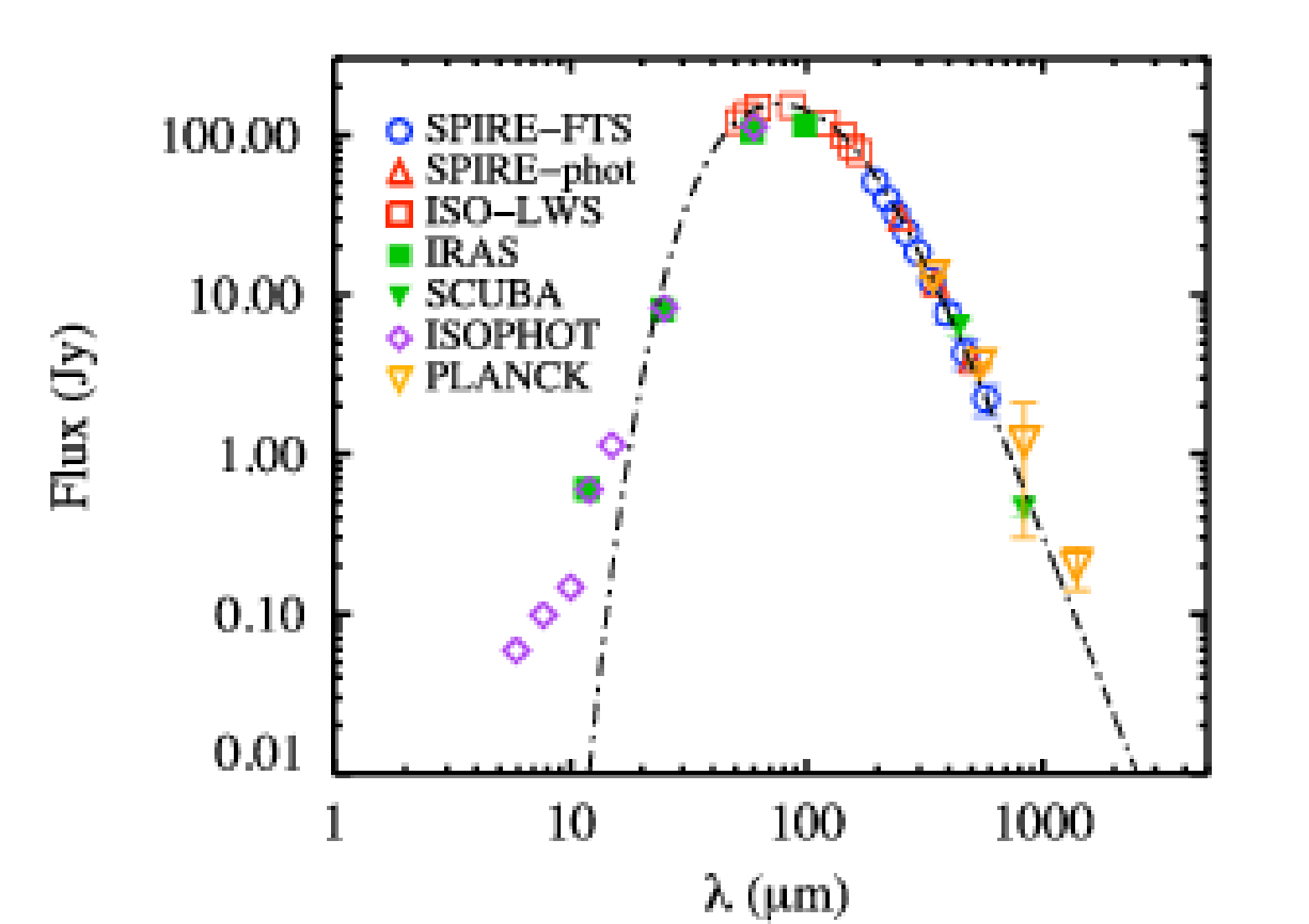}
\includegraphics[width=8.cm]{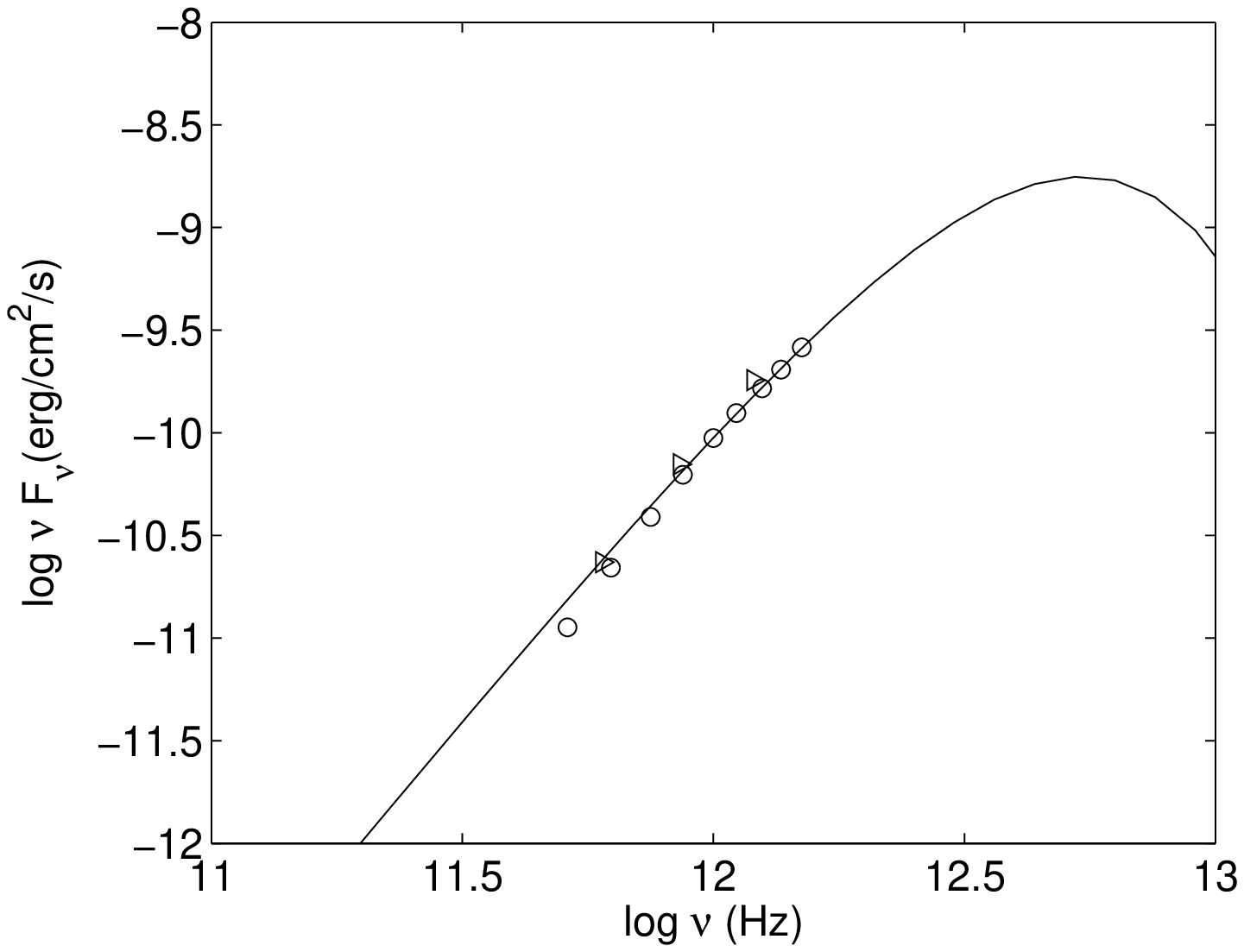}
\caption{Previous modelling of Arp.220. Left :
 Rangwala et al (2011, fig. 3); right : the slope of Rangwala et al.  data.
Black line : the Planck function corresponding to T=66.7 K.
}
\end{figure*}

\begin{figure*}
\includegraphics[width=14.0cm]{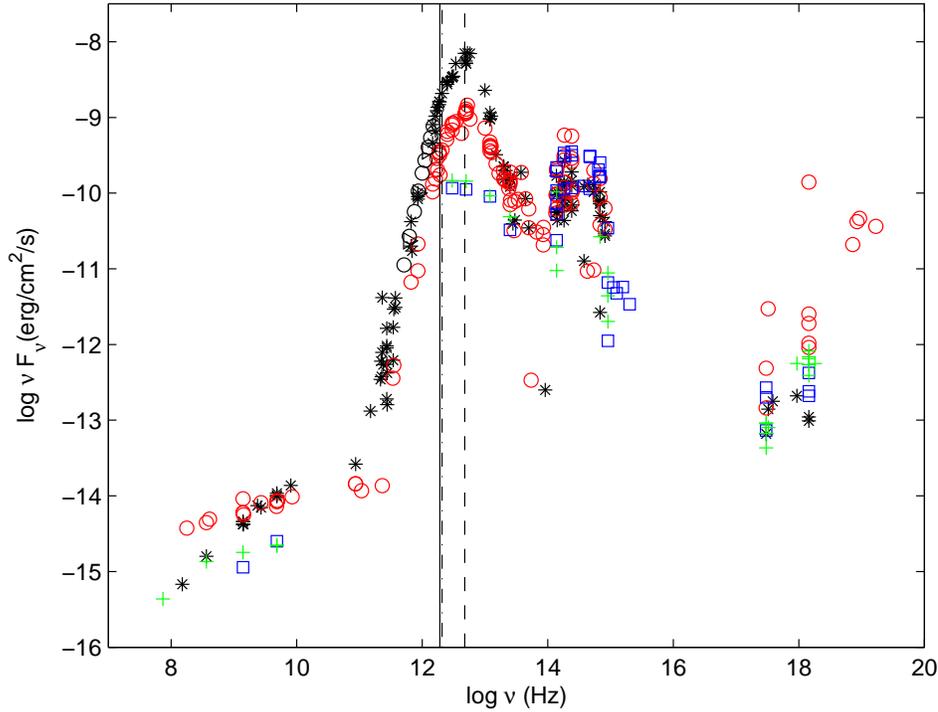}
\caption{
Black asterisks: Arp 220; red circles : NGC 6240; blue squares : NGC 3393;
green crosses : NGC 7212. Vertical lines : frequencies corresponding
to [CII] 157 (solid), [OI] 63 (dashed), [OI] 145 (dash-dotted).
}
\end{figure*}

\begin{figure*}
\includegraphics[width=14.0cm]{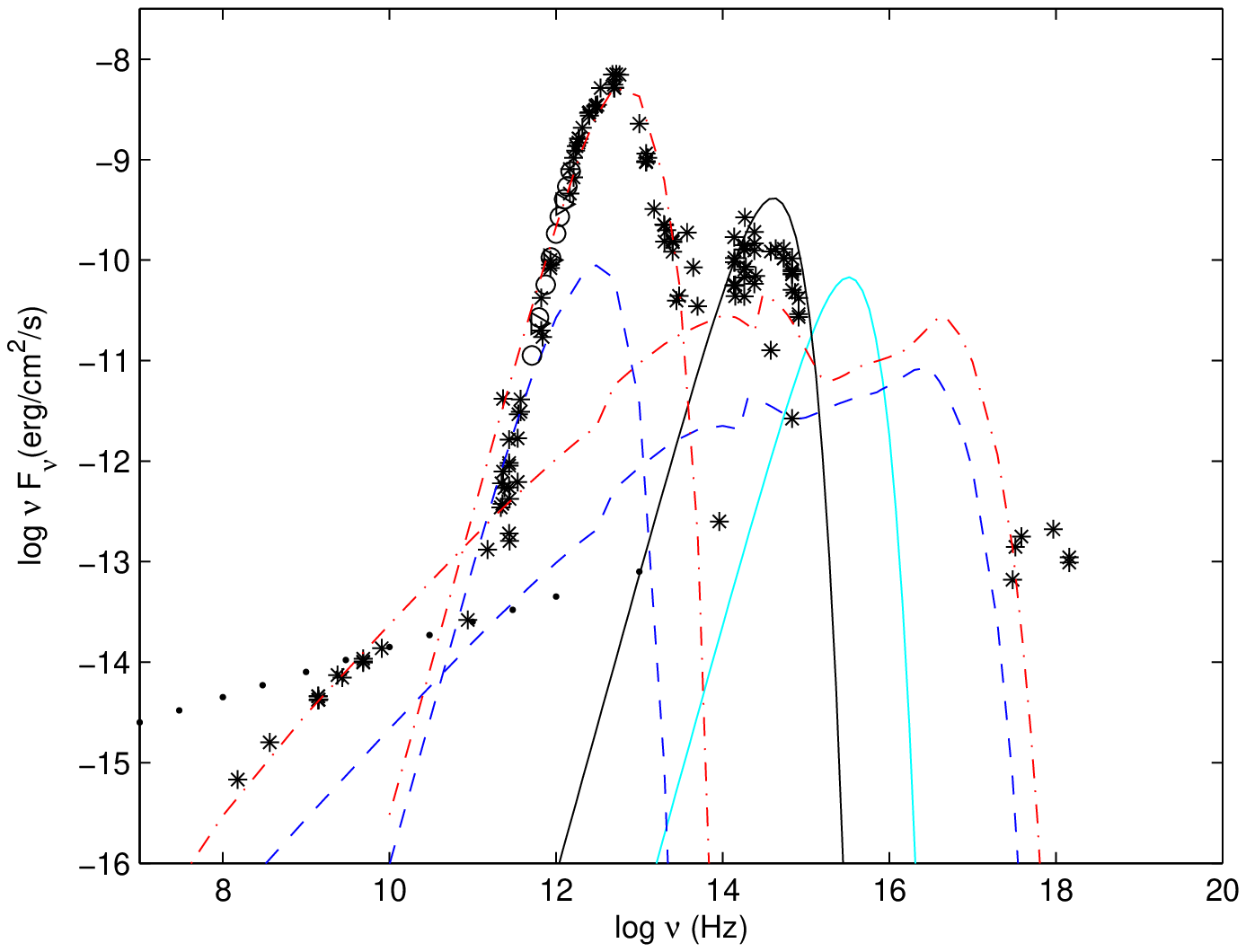}
\includegraphics[width=14.0cm]{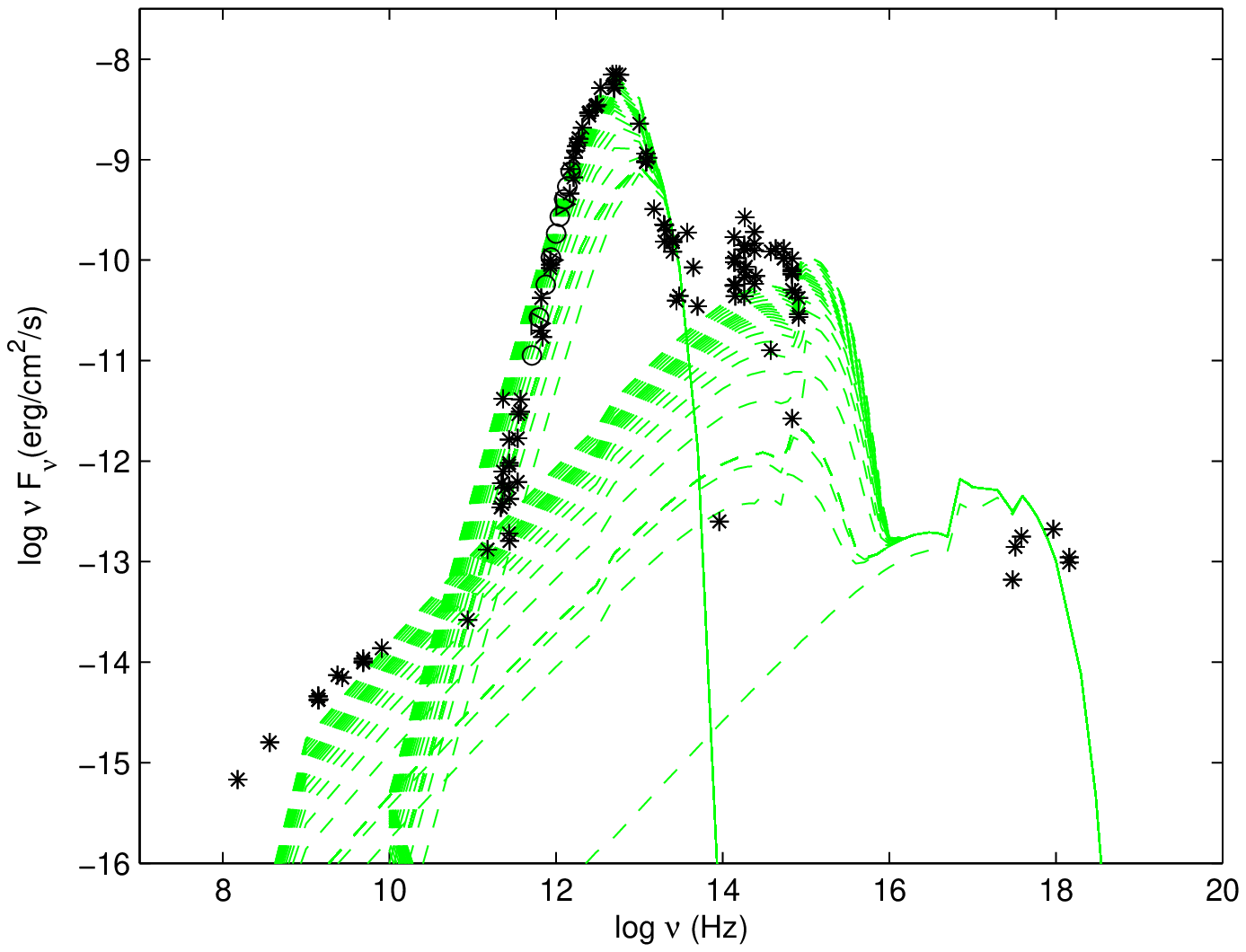}
\caption{
Top :
Black asterisks : the data from the NED;  black open circles and triangles : data from Rangwala et al (2011) by SPIRE-FTP and
SPIRE-photometer, respectively.
Blue dashed lines :  model  m1$_{pl}$ ;  red dot-dashed :  model m2$_{sb}$ ;
 black solid curve : the old star background bb flux  referring to T=4000 K; 
cyan solid line : bb flux  corresponding to \Ts=4 10$^4$ K;  dotted black line : synchrotron radiation.
Bottom :  green : gas and dust radiation from the slabs downstream for the model  corresponding to
 \Vs=1000 \kms, \n0=1000 \cm3 and log($F$)=11.
}
\end{figure*}

\section{The SED of the continuum}

Arp 220 continuum SED has been studied previously on a large frequency range  (e.g. Contini, Viegas \& Prieto
2004,  Contini \& Contini 2007) and in the IR domain (e.g. Rangwala et al. 2011).

We  report in Fig. 5  left and right panels the modelling of the SED based on a set of models which 
were selected investigating the IR bump 
of dust reprocessed radiation in  AGN (Contini et al 2004, fig. A1) and by
fitting the lines of luminous infrared galaxies by  Contini \& Contini (2007, fig. 6), respectively.
It can be noticed (Fig. 5, left)
that a model calculated by \Vs=1000 \kms, \n0=1000 \cm3,  a pl  flux F=10$^{11}$ cm$^{-2}$s$^{-1}$eV$^{-1}$
 at the Lyman limit and d/g=0.004, \agr=1 \mum, D=10$^{18}$ cm,
reproduces the whole dust reradiation bump, although two other models cannot be excluded (Fig. 5, right). 
One reproduces partially
the data at higher frequencies (\Vs=500 \kms, \n0=300 \cm3, U=10, \Ts=10$^4$ K, d/g=0.04, D=10$^{19}$ cm) 
and the other at lower frequencies  
(\Vs=150 \kms, \n0=100 \cm3, U=0.1, \Ts=10$^4$ K, d/g=3 10$^{-5}$,  D=10$^{15}$ cm).
A higher \Vs corresponds to a higher gas temperature downstream and, consequently, it yields a higher 
collisional heating of the grains by the gas.
The fit of the reradiation bump by different models evidences the fact that the  models cannot be constrained
only by the continuum SED.

In  Fig. 6  Rangwala et al (2011, fig. 3) is reproduced. Rangwala et al claim that  Arp 220 data in the IR peak can 
be  explained by the Planck function corresponding to one single temperature (T=66.7 K), as can be seen in detail
in the right  panel of Fig. 6. Indeed, Rangwala et al (2011, fig. 3)  reproduces 
the  IR peak of dust. Nevertheless, at  wavelengths $<$ 20 \mum the data  follow a different slope (Fig. 6, left panel).

In Fig. 7  we compare the data  of Arp 220 
 with the data presented for some other merger galaxies, NGC 3393, NGC 6240 and NGC 7212  which were previously
investigated by a detailed modelling.
We make use of the data from the NED for Arp 220
 (Ueda et al 2001,
Teng et al 2009,
    de Vaucouleurs et al (1991),
Violet et al (2005)
 Koulouridis (2006),
    Zwicky et al (1963),
Jason et al (2000),
Weedman et al (2009),
Klaas et al (2001),
 Gorjian et al (2004), 
 Soifer et al. (1989),
 Spinoglio  et al (2002),
Brauher et al. (2008),
 Klaas et al (2001),
Rigopoulou et al (1996)
 Eales et al (1989),
Matsushita et al (2009),
Dunne \& Eales (2001),
Anton et al (2004),
 Dunne (2000),
Sakamoto et al (2009),
Aalto et al (2009),
Carico et al (1992),
    Thronson et al (1987),
    Chini et al (1986),
Matsushita et al. (2009),
Anton et al (2004),
Imanishi et al (2007),
Condon et al. (1983)
Robert et al. (1991),
    Glen et al (1990),
    Condon et al (1983),
 Dressel et al (1978),
 Baan  \& Klockner (2006),
 Condon et al (1998),
   Condon et al (2002),
  White et al  (1992)
 Douglas et al 
Waldram et al (1996)
).  The black open  circles and triangles come from Rangwala et al (2011).

The data from the NED  were taken with different apertures and at different epochs,
however, we have shown by modelling the continuum of the NGC 7212  galaxy
(Contini et al 2012) that all the data are significant. In particular, the data
taken with  different large apertures are nested within the Plank function corresponding to the
old star background temperature.

Comparing with the merger galaxy NGC 6240,
an important issue about Arp 220 continuum appears.   
The radio synchrotron  radiation created by the Fermi mechanism at the shock front is  not seen.
 On the contrary, it is undoubtly present in  many other ULIRGs  
(Contini \& Contini 2007, fig. 6).
The most probable explication is  connected with the  large amount of dust in Arp 220.
Ginzburg \& Syrovatskii (1965) claim that the 
intensity and polarization of {\it magnetobremsstrahlung}  change following not only the source nature, but
the  nature of the medium  where it propagates from  source to Earth,
I=I$_0$ e$^{-{\tau}}$.  A sharp decrease of the intensity will follow at  low frequencies  
when $\tau$ becomes large.
Moreover, when  alternate regions of radiating and absorbing material (clouds) appear,
as it is suggested  by  high fragmentation in Arp 220,
there is a   strong decrease of the flux  at low frequencies.  
 The original spectral index of the observed flux at the source will change,  
and will become similar to that of the thermal bremsstrahlung
deleting any trace of synchrotron radiation.

In Fig. 7,   comparing the SED from the NED data of Arp 220 with the SED of
 NGC 6240, we notice
 that the IR reradiation bump peaks at the same frequency ($\sim$ 8 10$^{12}$ Hz,
but the intensity of the peak is higher for Arp 220 by about 1 order of magnitude, indicating a higher dust-to-gas ratio.
Both galaxies show  absorption at frequencies $\leq$ 10$^{12}$ Hz.
The old star black body  flux shows that the old star temperature  is  similar for the
two galaxies (4000 K).
In  Arp 220  the radio flux shows a steep slope, characteristic of  thermal bremsstrahlung.
It seems (presently) that Arp 220 is less luminous than NGC 6240 in the soft X-ray range.

We have traced upon the SED of the galaxies in Fig. 7 the frequencies corresponding to the [CII] 158,
[OI] 63 and 145 lines. Interestingly these lines appear on top of the dust reradiation peak.
The lines are emitted from the gas, while  the IR peak  from the dust. 
Fig. 7 suggests that at high $d/g$ ratios the far-IR lines can be underestimated (see also Contini 2011, fig.9). 

We would like to explain the SED of the continuum emitted from  Arp 220 by the same models
which were selected by the fit of  the line ratios. The observations are rich of data 
throughout the extended range of frequencies, from radio to X-ray.
In Fig. 8 (top panel)  we compare the SED of the models  with the data.
Each model corresponds to two lines, one is the bremsstrahlung from the gas and the other is the dust 
reprocessed radiation flux.
Model m1$_{pl}$ nicely fits the radio bremsstrahlung slope and the soft X-rays (blue lines).
Model m2$_{sb}$  (red lines) is calculated by  
a grain radius \agr=0.30 \mum and $d/g$=0.02 by mass. It overpredicts the low frequency side of the
dust reradiation bump, indicating absorption at $\nu$$\leq$ 10$^{12}$ Hz.

We  show in Fig. 8 (bottom panel) the model calculated to fit the  Arp 220 SED 
in a previous paper (Contini et al 2004) with a high \Vs =1000 \kms , a high \n0 = 1000 \cm3, and a large \agr=1 \mum.
Those high velocities were suggested also by the broad Br$\alpha$ reported by DePoy et al (1987).
The model calculated with log($F$)=11  reproduces satisfactorily the whole dust reradiation bump,
explains the  X-ray data and the data in the near-IR.

Recall that the models are calculated at the nebula while the data are observed at Earth.
So the models are shifted on the Y-axis adopting relative weights which 
depend on (R/d)$^2$, where R is the distance of the emitting clouds from the AC and d the distance from Earth.
 The radiation dominated high velocity model is shown in Fig. 8  adopting a very low (0.01)
 weight relative to the models explaining the line ratios (m1$_{pl}$ and m2$_{sb}$).

The shock dominated model  referring to \Vs=1000 \kms 
yields acceptable IR line ratios (Table 2, model m$_{sd}$), while
the high velocity  model dominated by the radiation flux from the AGN leads to  high [OIII]/\Hb. However, these lines 
(which are not shown in Table 1 to avoid confusion)
do not spoil the good fit  to the data because they should be added to the lines emitted from the
other models which appear in Table 1 with the same low relative weight used to fit the SED.

In particular, we show in  Fig. 8 (bottom panel)  the proceeding  of the SED calculation by 
summing up the different slab
contributions downstream (not all are shown in the figure for sake of clarity).
The final SED profile  results from the integration
of the single slab emissions. In the IR peak, the  continuum radiation from  each single slab  
results from the Planck function
corresponding to the temperature calculated in that slab. Actually, this temperature derives from the
mutual heating and cooling of dust and gas.

\section{Discussion and concluding remarks}

\begin{table*}
\begin{center}
\caption{Comparison of physical conditions in different merger galaxies}
\small{
\begin{tabular}{lccccccccccc} \hline  \hline
\ galaxy      &z        &model &    \Vs  & \n0  & $F$  & $U$  & \Ts  &   $D$ & \agr & $d/g$   \\
\             &         &    & \kms      & \cm3 & units$^1$&-  &   K  &   cm & \mum & by mass   \\ \hline
\ NGC 3393 $^2$&  0.0125&pl &    100     & 3000 & 2.3e12&-    & -    & 4.9e16& 1.& 1.6e-4  \\
\             &         & bb &    600    & 300  &  -    &1.   & 8.6e4& 1.e17 &  1. &2.e-4    \\
\             &         & sd &    300    & 1500 &  -    & -   &  -   &  1.e16&  1. & 0.004   \\
\             &         & sd &  1000     & 1500 & -     & -   & -    & 3.e18 & 1.& 0.004\\
\ NGC 6240 $^3$& 0.0245&  pl & 500-900   & 40   &  3.e8 & -   & -    & 6.e17 &  1. & 0.004  \\
\             &         & bb & 500       & 68   &  -    & 0.002& 5.e4& 1.e19 &  1. & 4.e-4 \\
\             &         & sd & 2000      & 500  & -     &  -   & -    &  1.e19 & 0.5& 0.004 \\
\ Arp 220 $^4$ & 0.018& pl   & 300       & 30   & 2.5e8 & -   &   -  & 1.5e17& 1.2 & 4.e-6  \\
\             &         &bb  & 400       & 40  & -     &$<$0.008 &  4.5e4 & 3.e18 & 0.30&0.02  \\
\             &        & sd  &1000       &1000 & (1.e11)&-    &  -   & 1.e17 & 1. & 4.e-4  \\
\ NGC 7212 $^5$&0.0266 & pl  &110-550    &70-120& $\sim$ 7.e10 &-&-  & 2.e17 & 1. & 4.e-4 \\ \hline

\end{tabular}

$^1$ in photon cm$^{-2}$ s$^{-1}$ eV$^{-1}$ at the Lyman limit

Ages : $^2$ $\sim$ 5 Myr (Mayer et al 2007) ;$^3$ $<$5 Myr (Engel et al 2010); $^4$10-300 Myr   (Wilson et al 2005);

$^5$100 -1000 Myr (Ramos Almeida et al 2009)
}
\end{center}
\end{table*}

\begin{table*}
\begin{center}
\caption{Comparison of element relative abundances in different merger galaxies}
\begin{tabular}{lcccccccc} \hline  \hline
\ galaxy   & mod& C/H  & N/H           & O/H          & Si/H  & S/H    \\
\         & &  10$^{-4}$&10$^{-4}$ &10$^{-4}$&10$^{-5}$&10$^{-5}$    \\ \hline
\ NGC 3393&pl$^1$& 2.3& 1.8         & 6.6  &3.3     &1.6    \\
\         &bb& 3.3&1.8        &  6.6 &3.3      & 1.6          \\
\         &sd& 3.3 & 1.8      &  6.6 &3.3      & 1.6  \\
\         &sd& 3.3 & 0.91     & 6.6  & 3.3  & 1.6    \\
\ NGC 6240&pl& 4.3& 1.5        & 8.6 &3.3       &1.6     \\
\         &bb& 4.5&1.5        & 9.6  &2.      &0.4   \\
\         &sd& 4.0& 1.5       & 9.6  &3.3      & 1.6 \\
\ Arp 220 &pl& 3.3& 1.4       & 3.6  &0.2      &1.1 \\
\         &bb& 3.3& 1.5       & 9.6  &0.33       &1.0  \\
\         &sd&  3.3&0.91       &6.6  &3.3       & 0.8 \\
\ NGC 7212&pl& 3.3 &0.3-2.3 &2.6-6.6 &3.3&0.37-2.8   \\ \hline

\end{tabular}

$^1$ Mg/H = 8.5e-6

\end{center}
\end{table*}

The physical conditions and the relative abundances of heavy elements to H have been calculated
by modelling the line and continuum spectra  of Arp 220 in the optical and IR ranges.
 In Tables 3 and 4 we compare the results for Arp 220 with those obtained by the detailed
modelling of other merger galaxies (NGC 3393, Contini 2012a, NGC 7212, Contini et al 2012, NGC 6240, Contini 2012b). 
The ages of the galaxies are given in the bottom of Table 3.
The models which appear in Table 3, column 3, are characterised  by the nature of the radiation
flux, namely a power-law (pl), black body (bb), and shock dominated (sd)  which indicates that $F$=0 or $U$=0.  

\subsection{Results for Arp 220}

Although the line spectra of Arp 220 in the different frequency ranges are given by  single
observations
covering all the different regions within the extended NLR,
we can at least try to find  the location of the clouds  in the surroundings of the AGN and  of the starburst. 
First notice that the  clouds  emitting the spectra show similar shock velocities and preshock densities.
The   preshock densities are similar to those calculated in the Milky Way ISM.
 The optical spectra  observed from Arp 220 come from clouds reached by the AGN
radiation flux, while the IR spectra are emitted from clouds close to the starburst.
It seems that the strong obscuration in the  galaxy central region is  fragmented
such as to allow  to observe the optical spectrum from the NLR of the AGN.

The distance R of the emitting gas from the starburst
  and from the AGN, can be calculated
combining the observed absolute flux observed at Earth with the  absolute flux
calculated at the nebula F$_{\lambda}$ (obs) d$^2$ =F$_{\lambda}$ (calc) R$^2$ $W$, 
where d is the distance from Earth and $W$ can be regarded as the filling factor.
For Arp 220  d=72 Mpc is adopted.

We refer to the absolute values of the  [NeII]12.8, [CII] 158 and  \Hb lines as representative
of the IR, far-IR and optical ranges, respectively.
The results show  that the clouds emitting the mid-IR  and the far-IR  lines, e.g. [NeII]12.8 and [CII]157, 
respectively, are at a distance R$_{[NeII]}$=2.96 kpc 
and R$_{[CII]}$=5.64 kpc from the starburst, adopting $W$=1.    
In NGC 6240 the starburst fills the central kpc region encompassing the two nuclei (Engel et al. 2010).
If the 4.5 10$^4$ K stars  embody the central starburst of Arp 220, the clouds  emitting  the  IR and FIR
lines are close to  it, therefore they are highly obscured in the optical range.

We have shown in Sect. 3.3 that the  [OI]63, 145 and [CI] at 492 and 809 GHz lines are formed in the internal region 
of extended clouds and are therefore absorbed, while [CII] lines are emitted from the external edges.
This confirms that the clouds are moving outwards.
In fact, the temperatures on the edge of the cloud facing the radiation source reach a maximum
of 2-3 10$^4$ K because the gas is heated by radiation. On the opposite edge, the gas is heated collisionally
by the shock at temperatures which depend on \Vs. Figs. 1 and 2 show that the temperatures on the shock front
are $\sim$ 10$^6$ K yielding to highly ionized gas. Such temperatures are certainly higher 
than those corresponding to   the first ionization level lines.
On the contrary, in the inflowing case, radiation from the external ionizing source would reach the clouds on the
shock front,  where collisional ionization dominates. Consequently  the [CII] line emission will be
restricted to a small region  in the internal region of the clouds.

 R$_{H{\beta}}$=1.51 kpc
is the distance from the active centre of the optical emitting clouds, adopting  $W$=0.01 (Sect. 3.3). 
The optical emission  comes from the extended NLR of the  AGN,  where the extinction is relatively  low.
The O/H ratio is lower than solar by a factor of $\sim$ 2 and  S/H
by a small factor of 1.4. We suggest that  such low abundances are characteristic of matter included
in  the merger during the parent collision.
The high velocity clouds are located at a distance of 66 kpc from the active centre.

The relative abundances (Table 3)   calculated for the Arp 220 starburst show that   O/H are higher than  solar, 
while Si /H and
S/H highly depleted, indicating inclusions into dust grains throughout the starburst region.
A high dust-to-gas ratio is found by modelling the dust reprocessed radiation bump by  the model which
represents the clouds  photoionised by the starburst.
This could explain also the  strong absorption of the synchrotron radiation created by the Fermi mechanism 
at the shock front which
 characterizes  the Arp 220  radio SED. Only the
radio thermal bremsstrahlung is  seen in the continuum SED.

We have found that the  dust reprocessed radiation bump in the IR is well 
reproduced by relatively large grains
(\agr=1 \mum) accompanying the gas throughout the shock front and downstream for \Vs=1000 \kms.
Grains with \agr=0.3 \mum  are adopted in the model which represents clouds close to the starburst.
They also yield a satisfactory fit of the IR bump but do not cover the X-ray domain. 
 Grain growth up to cm-sized  particles  are predicted by Ubach et al (2012).
 So grains of $\sim$ 1 \mum size entering the shock front
 could  be the product of former larger grain  fragmentation.

\subsection{Comparison with other merger AGN}

We compare now  the results of the physical conditions and of the relative abundances of the heavy elements, 
obtained  for Arp 220 in previous sections with the results obtained by modelling other 
merger galaxies.
Two AGNs were observed in the X-ray in  NGC 3393  (Fabbiano et al 2011) and in
NGC 6240 (Komossa et al 2003).
In NGC 7212 the merging  is  confirmed by patches of gas  emitting broad line profiles at the edges of the ENLR
(Cracco et al. 2011).

Table 3 shows that
the shock velocities are similar  to those generally found in the NLR of AGN (100-1000 \kms). 
However, it is important for mergers to investigate
the distribution of \Vs throughout the NLR in order to find out  eventual collision records.
The shock velocities are between 300 and 1000 \kms, except for  NGC 3393 and NGC 7212 where 
regions of low velocity
survived  to the galaxy collision throughout most of the NLR. The relatively high \Vs observed  in some central  regions 
and at the outer limits of the  extended NLR
shows  residuals  of collision. The shocked matter seems present throughout most of the NGC 3393 NLR, 
not only because of
the high \Vs but   because of the  unexpectedly high \n0  close to the AGN.
In the other galaxies presented in Table 3, the
 preshock densities range from 30-40-120 \cm3 in NGC 6240,  Arp 220 and NGC 7212, respectively.

A starburst was found in the centre of the merger galaxies, except in NGC 7212 where  the eventual 
double AGN and the central starburst is  not yet observed.

Relative abundances (Table 4)  show generally a higher than solar N/H, by a factor of $\sim$1.5.
except for NGC 7212 showing N/H depletion in most of the observed  regions. 
In particular Arp 220 shows N/O  $\sim$ 0.2-0.4, higher than the solar value N/O=0.15 
(Allen 1976)   and N/O=0.12  (Anders \& Grevesse (1989).
This result can be explained following Contini et al (2002b), namely  the galaxy ISM
 can be enriched in nitrogen but not in oxygen during a long period of quiescence of intermediate-mass 
star evolution.

The results obtained by modelling the Arp 220 line and continuum spectra are only a hint to the real picture
of the physical conditions throughout the galaxy. There is an urgent need of integral spectroscopy
in order to have more information  about the distribution of the physical and chemical quantities.

\section*{Aknowledgements}
I am very grateful to the referee for important remarks which led to a more correct
presentation  of the results.
I wish to thank Dr. N. Rangwala for allowing to reproduce fig.3 of Rangwala et al.
(2011).
This research has made use of the NASA Astrophysics Data System (ADS) and the NE
D, which is operated
by the Jet Propulsion Laboratory, California Institute of Technology, under contract with NASA.

\section*{References}

\def\ref{\par\noindent\hangindent 20pt}

\ref Aalto, S.; Wilner, D.; Spaans, M.; Wiedner, M. C.; Sakamoto, K.; Black, J.
H.; Caldas, M.  2009,A\&A,493,481
\ref Allen, C.W., 1976, Astrophysical Quantities, London: Athlone (3rd edition)
\ref Anders, E., Grevesse, N. 1989, Geochin. Cosmochim. Acta, 53, 197
\ref Anton, S.; Browne, I. W. A.; Marcha, M. J. M.; Bondi, M.; Polatidis, A.  2004,MNRAS,352,673
\ref Armus, L., Heckman, T.M., Miley, G.K. 1989, ApJ, 347, 727
\ref Armus, L. et al 2006, ApJ, 640, 210
\ref  Baan, W.A.,   Klockner, H. -R.  2006, A\&A,449,559

\ref Becker, R.H.,  White, R.L.,  Edwards, A.L.  1991,ApJS,75,1

\ref Brauher, J. R.; Dale, D. A.; Helou, G. 2008, ApJS,178,280

\ref Carico, D.P. ,  Keene, J.,  Soifer, B.T.,  Neugebauer, G.  1992, PASP,104,1086

\ref Chini R., Kreysa E., Krugel E., Metger P.G.  1986. A\&A, 166,  L8 

\ref Condon, J.J.,  Cotton, W.D.,  Greisen, E.W.,  Yin, Q.F.,  Perley, R.A.  Taylor, J.B.,  Broderick, J.J.  1998,AJ,115,1693

\ref Condon, J.J.,  Cotton, W.D.,  Broderick, J.J. 2002,AJ,124,675

\ref Condon, J.J., Condon, M.A., Broderick, J.J., Davis, M.M.  1983, AJ,  88,  20

\ref Contini, M. 2012a, MNRAS, 452, 1205
\ref Contini, M. 2012b, MNRAS, 426, 719
\ref Contini, M., Cracco, V., Ciroi, S., La Mura, G. 2012, A\&A, 545, 72 
\ref Contini, M., Goldman, I. 2011, MNRAS, 411, 792
\ref Contini, M. 2009, MNRAS, 399, 1175
\ref Contini, M. 2011 MNRAS, 418, 193
\ref Contini, M. 2004, MNRAS, 348, 1065
\ref Contini, M., Radovich, M., Rafanelli, P., Richter, G.M. 2002a, ApJ, 572. 124 
\ref Contini, M., Contini, T. 2007, AN, 328, 953
\ref Contini, M.; Viegas, S. M.; Prieto, M. A.  2004, MNRAS, 348, 1065
\ref Contini, M. 1997, A\&A, 323, 71
\ref Contini, M., Aldrovandi, S.M. 1983, A\&A, 127, 15
\ref Contini, T., Treyer, M.~A., Sullivan, M., \& Ellis, R.~S. 2002b, MNRAS, 330, 75
\ref Cracco, V. et al. 2011, MNRAS, 418, 2630
\ref De Vaucouleurs, G., De Vaucouleurs, A., Corwin JR., H.G., Buta, R. J. Paturel, G., and Fouque, P.
    Third Reference Catalogue of Bright Galaxies, version 3.9 1991 vol. p.  RC3.9

\ref DePoy, D.L., Becklin, E.E., Geballe, T.R. 1987, ApJ, 316, 63
\ref Douglas, J.N.,  Bash, F.N.,  Bozyan, F.A.,  Torrence, G.W.  Dressel, L.L.,  Condon, J.J. 1978,ApJS,36,53
\ref Draine, B.T. 2009, in Interstellar Dust from Astronomical Observations to
Fundamental Studies, F. Boulanger, C. Joblin, A. Jones, and S. Madden (eds),
EAS Publication Series, 35, 245,268

\ref Dunne, L. and  Eales, S.A.  2001, MNRAS, 327, 697

\ref Dunne, L.,  Eales, S.,  Edmunds, M.,  Ivison, R.,  Alexander, P.,  Clements, D.L. 2000, MNRAS, 315,15
\ref  Eales, S.A., Wynn-Williams, C.G.,  Duncan, W.D. 1989,ApJ,339,859
\ref Emerson, J.P., Clegg, P.E., Gee, G., Griffin, M.J., Cunningham, C.T., Brown, L.M.J., 
Robson, E.I., Longmore, A.J. 1984,Natur, 311,237
\ref Engel, H. et al. 2010, A\&A, 524, 56 
\ref Farrah, D. et al 2007, ApJ, 667, 149
\ref Flower. D.R., Launay, J.M. 1977, JPhB, 10,3673
\ref Ginzburg, V.L.,   Syrovatskii, S.L. 1965 ARA\&A, 3, 297
\ref Gonzalez-Alfonso, E., Smith, H.A., Fischer, J., Cernicharo, J. 2004, ApJ, 613, 261
\ref Gorjian, V.,  Werner, M.W.,  Jarrett, T.H.,  Cole, D.M., and  Ressler, M.E.  2004,ApJ,605,156
\ref Graham, G.R., Carico, D.P., Matthews, K., Neugebauer, G., Soifer, B.T., Wilson, T.D. 1990, ApJ, 354, L5 

\ref Hayes, M.A. , Nussbaumer, H. 1984 A\&A, 139, 233 
\ref Imanishi, M.; Nakanishi, K.; Tamura, Y.; Oi, N.; Kohno, K. 2007,AJ,134,2366

\ref Jason A. Surace D. B. Sanders A. S. Evans 2000,ApJ,529,170

\ref Klaas, U.  et al.  2001,A\&A,379,823
\ref Komossa, S., Burwitz, V., Hasinger, G., Predehl, P., Kaastra, J.S., Ikebe, Y. 2003, ApJ, 582, L15

\ref Koulouridis, E.,  Plionis, M.,  Chavushyan, V.,  Dultzin-Hacyan, D.,  Krongold, Y.,  Goudis, C 2006,ApJ,639,37

\ref Langston, G.I.,  Helfin,M.B.,  Conner, S.R.,  Lehar, J.,   Carrilli, C.L.,  Burke, B.F. 1990,ApJS,72,621

\ref Luhman, M.L. et al 1998, ApJ, 504, L11
\ref Luhman, M.L., Satyapal, S., Fischer, J., Wolfire, M.G., Sturm, E., 
Dudley, C.C., Lutz, D., Genzel, R. 2003, ApJ, 594, 758
\ref Malhotra, S. et al 1997 ApJ, 491, 27
\ref Matsushita, S. et al.  2009,ApJ,693,56

\ref Osterbrock, D.E. 1989 in Astrophysics of gaseous nebulae and active galactic nuclei/ University Science Books, 1989
\ref Rangwala, N. et al. 2011 ApJ, 743, 94
\ref Rieke, G.H., Cutri, R.M., Black, J.H. et al. 1985, ApJ 290, 116
\ref Rigopoulou, D.,  Lawrence, A.,   Rowan-Robinson, M. 1996,MNRAS,278,1049

\ref Sakamoto, K. et al.  2009,ApJ,700,L104
\ref Simpson, J.P.; Colgan, S. W. J.; Cotera, A. S.; Erickson, E. F.; Hollenbach, D. J.; 
Kaufman, M. J.; Rubin, R. H. 2007, ApJ, 670, 111
\ref Smith, C.H., Aitken, D.K., Roche, P.F. . 1989, MNRAS, 241, 425

\ref Soifer, B.T.,  Boehmer, L.,  Neugebauer, G., and  Sanders, D.B.  1989,AJ,98,766

\ref Spinoglio, L., Andreani, P.,  Malkan, M.A. 2002,ApJ,572,105
\ref 	Stacey, G. J.; Geis, N.; Genzel, R.; Lugten, J. B.; Poglitsch, A.; Sternberg, A.; Townes, C. H.
1991, ApJ, 373, 423
\ref Stutzki, J. et al. 1997, ApJ, 477, L33

\ref Taylor, V.A.,  Jansen, R.A.,  Windhorst, R.A., . Odewahn, S.C.,  Hibbard, J.E. 
2005, ApJ, 630, 784
\ref Teng, S.H. et al.  2009, ApJ, 691, 261
\ref Thronson H.A., JR., Walker C.K., Walker C.E., Maloney P.
     1987, ApJ, 318, 645
\ref Ubach, C., Maddison, S.T., Wright, C.M., Wilner, D.J., Lommen, D.J.P. Koribalski, B.
2012arXiv1207,0260
\ref  Ueda, Y.,  Ishisaki, Y.,  Takahashi, T.,  Makishima, K., and  Ohashi, T.  2001,ApJS,133,1

\ref Veilleux, S., Kim, D.-C., Sanders, D.B. 1999, ApJ, 522,113 
\ref Viegas, S.M., Contini, M. 1994, ApJ, 428,113
\ref Waldram, E.M.,  Yates, J.A.,  Riley, J.M. and  Warner, P.J.  1996,AJ,111,1945
\ref Weedman, D.W.; Houck, J.R.  2009,ApJ,698,1682
\ref White, R.L.,   Becker, R.H. 1992, ApJS, 79, 331 
\ref Wilson, C.D., Harris, W.E., Longden, R., Scoville, N.Z. 2006, ApJ, 641, 763  
\ref Yasuda, A., Nakagawa, T., Spaans, M., Okada, Y., Kaneda, H. 2008, A\&A, 480, 157
\ref Zwicky, F., Herzog, E.
    "Catalogue of Galaxies and of Clusters of Galaxies", 1963, volume II, Pasadena: California Institute of Technology 1963 vol. p.

\end{document}